
%
\input phyzzx
\mathchardef\bftheta="0912

%
\catcode`@=11
%
%
\font\fourteenmib=cmmib10 scaled\magstep2   \skewchar\fourteenmib='177
\font\twelvemib=cmmib10 scaled\magstep1     \skewchar\twelvemib='177
\font\elevenmib=cmmib10 scaled\magstephalf  \skewchar\elevenmib='177
\font\tenmib=cmmib10                        \skewchar\tenmib='177
%
\font\fourteenbsy=cmbsy10 scaled\magstep2   \skewchar\fourteenbsy='60
\font\twelvebsy=cmbsy10 scaled\magstep1     \skewchar\twelvebsy='60
\font\elevenbsy=cmbsy10 scaled\magstephalf  \skewchar\elevenbsy='60
\font\tenbsy=cmbsy10                        \skewchar\tenbsy='60
%
%
\newfam\mibfam
%
%
\def\fourteenf@nts{\relax
    \textfont0=\fourteenrm          \scriptfont0=\tenrm
      \scriptscriptfont0=\sevenrm
    \textfont1=\fourteeni           \scriptfont1=\teni
      \scriptscriptfont1=\seveni
    \textfont2=\fourteensy          \scriptfont2=\tensy
      \scriptscriptfont2=\sevensy
    \textfont3=\fourteenex          \scriptfont3=\twelveex
      \scriptscriptfont3=\tenex
    \textfont\itfam=\fourteenit     \scriptfont\itfam=\tenit
    \textfont\slfam=\fourteensl     \scriptfont\slfam=\tensl
    \textfont\bffam=\fourteenbf     \scriptfont\bffam=\tenbf
      \scriptscriptfont\bffam=\sevenbf
    \textfont\ttfam=\fourteentt
    \textfont\cpfam=\fourteencp
    \textfont\mibfam=\fourteenmib   \scriptfont\mibfam=\tenmib
    \scriptscriptfont\mibfam=\tenmib }
\def\twelvef@nts{\relax
    \textfont0=\twelverm          \scriptfont0=\ninerm
      \scriptscriptfont0=\sixrm
    \textfont1=\twelvei           \scriptfont1=\ninei
      \scriptscriptfont1=\sixi
    \textfont2=\twelvesy           \scriptfont2=\ninesy
      \scriptscriptfont2=\sixsy
    \textfont3=\twelveex          \scriptfont3=\tenex
      \scriptscriptfont3=\tenex
    \textfont\itfam=\twelveit     \scriptfont\itfam=\nineit
    \textfont\slfam=\twelvesl     \scriptfont\slfam=\ninesl
    \textfont\bffam=\twelvebf     \scriptfont\bffam=\ninebf
      \scriptscriptfont\bffam=\sixbf
    \textfont\ttfam=\twelvett
    \textfont\cpfam=\twelvecp
    \textfont\mibfam=\twelvemib   \scriptfont\mibfam=\tenmib
    \scriptscriptfont\mibfam=\tenmib }
\def\tenf@nts{\relax
    \textfont0=\tenrm          \scriptfont0=\sevenrm
      \scriptscriptfont0=\fiverm
    \textfont1=\teni           \scriptfont1=\seveni
      \scriptscriptfont1=\fivei
    \textfont2=\tensy          \scriptfont2=\sevensy
      \scriptscriptfont2=\fivesy
    \textfont3=\tenex          \scriptfont3=\tenex
      \scriptscriptfont3=\tenex
    \textfont\itfam=\tenit     \scriptfont\itfam=\seveni  
    \textfont\slfam=\tensl     \scriptfont\slfam=\sevenrm 
    \textfont\bffam=\tenbf     \scriptfont\bffam=\sevenbf
      \scriptscriptfont\bffam=\fivebf
    \textfont\ttfam=\tentt
    \textfont\cpfam=\tencp
    \textfont\mibfam=\tenmib   \scriptfont\mibfam=\tenmib
    \scriptscriptfont\mibfam=\tenmib }
\def\mib{\n@expand\f@m\mibfam}

%
%
\Twelvepoint
\catcode`@=12
%

%
\def\tr{\mathop{\rm tr}\nolimits}
\def\deltaGam{\delta'_\Gamma}  
\def\Kex{{\bf K}}              
\def\Apara{{\cal A}}           %
\def\Gglobal{G_{\rm global}}
\def\Hlocal{H_{\rm local}}
\def\Hdiag{H_{\rm diag}}
\def\Klinear{\Kex\hbox{-}{\rm linear}}
\def\Kquadratic{\Kex\hbox{-}{\rm quadratic}}
\def\VA{{\mib V}}
\def\dBRS{\delta_{\rm B}}
\def\Apm{{\cal A}_{\mu\perp}}
\def\alpm{\alpha_{\mu\perp}}
\def\GI{{\rm GI}}
\def\a{\alpha}
\def\gp{{g^{\perp}_{b\alpha}}^a}
\def\gampla{\gamma^{\perp}_{l[\alpha]}}
\def\ttgamp{\tilde{\tilde\gamma}^{\perp}_{l\nu_1}}

\def\ee{\eqno\eq}
\def\para{\parallel}
\def\mtil{\tilde{m}}
\def\Hgf{V_\mu}               
\def\tilde{\widetilde}      
\def\hat{\widehat}          
\NPrefs
\def\PR#1#2#3{{ Phys. Rev. }{\bf #1} {(#3)} #2}
\def\PRL#1#2#3{{ Phys. Rev. Lett. }{\bf #1} {(#3)} #2}
\def\PL#1#2#3{{ Phys. Lett. }{\bf #1} {(#3)} #2}
\def\Physica#1#2#3{{ Physica }{\bf #1} {(#3)} #2}
\def\AP#1#2#3{{ Ann. of Phys. }{\bf #1} {(#3)} #2}

\def\NP#1#2#3{{ Nucl. Phys. }{\bf #1} {(#3)} #2}
\def\PREP#1#2#3{{ Phys. Rep. }{\bf #1} {(#3)} #2}
\def\PROG#1#2#3{{ Prog. Theor. Phys. }{\bf #1} {(#3)} #2}

\REF{\BandoKugoYamawaki}{
M.~Bando, T.~Kugo and K.~Yamawaki, \PREP{164}{217}{1988}.}
\REF{\BandoKugoUeharaYamawakiYanagida}{
M.~Bando, T.~Kugo, S.~Uehara, K.~Yamawaki and T.~Yanagida,
\PRL{54}{1215}{1985}.}
\REF{\Sakurai}{
J.J.~Sakurai, {\it Currents and mesons}
(Chicago Univ. Press, Chicago, 1969).}
\REF{\KSRF}{
K.~Kawarabayashi and M.~Suzuki, \PRL{16}{255}{1966};
Riazuddin and Fayyazuddin, \PR{147}{1071}{1966}.}
\REF{\BandoKugoYamawakiNP}{
M.~Bando, T.~Kugo and K.~Yamawaki, %
\NP{B259}{493}{1985}.}
\REF{\BandoKugoYamawakiPTP}{
M.~Bando, T.~Kugo and K.~Yamawaki, %
\PROG{73}{1541}{1985}.}
\REF{\HaradaYamawaki}{
M.~Harada and K.~Yamawaki,
\PL{B297}{151}{1992}.}
\REF{\HaradaKugoYamawakiPRL}{
M.~Harada, T.~Kugo and K.~Yamawaki,
DPNU-93-01/KUNS-1178.}
\REF{\BRS}{
C.~Becchi, A.~Rouet and R.~Stora,
\AP{{\rm(NY)}98}{287}{1976}.}
\REF{\refZinn}{
J.~Zinn-Justin, ``Renormalization of Gauge Theories''
in {\it Lecture Notes in Physics} {\bf 37}
(Springer, 1975);
B.W.~Lee, ``Gauge Theories'' in
{\it Methods in Field Theory},
Les Houches 1975
(North-Holland, 1976).}
\REF{\BlasiCollina}{
A.~Blasi and R.~Collina,
\NP{B258}{204}{1987};\PL{B200}{98}{1988}.}
\REF{\refBecchi}{
C.~Becchi, A.~Blasi, G.~Bonneau, R.~Collina and F.~Delduc,
Commun. Math. Phys. {\bf 120} (1988) 121.}
\REF{\refCCWZ}{
C.G.~Callan, S.~Coleman, J.~Wess and B.~Zumino,
\PR{177}{2247}{1969}.}
\REF{\refBB}{
M.~Boulware and L.S.~Brown, \AP{{\rm (N.Y.)}138}{392}{1982}.}
\REF{\refAppBer}{
T,~Appelquist and C.~Bernard, \PR{D23}{425}{1981}.}
\REF{\refWeinbergChPT}{
S.~Weinberg, \Physica{96A}{327}{1979}.}
\REF{\refNakanishi}{
N.~Nakanishi, \PROG{51}{952}{1974}.}
\REF{\FKTUY}{
T.~Fujiwara, T.~Kugo, H.~Terao, S.~Uehara and K.~Yamawaki,
\PROG{73}{926}{1985}.
}
\REF{\Georgi}{
H.~Georgi, \PRL{63}{1917}{1989};
\NP{B331}{311}{1990}.}
\REF{\WeinbergRho}{
S.~Weinberg, \PR{166}{1568}{1968}.}

\date{%
KUNS-1179 \cr
HE(TH) 93/02\cr
DPNU-93-02\cr
March, 1993}

\titlepage
\title{
Low Energy Theorems of Hidden Local Symmetries
}
\author{
Masayasu Harada\foot{
Address after April 1, 1993 :
Department of Physics, Kyoto University, Kyoto 606-01, Japan
}}
\address{
Department of Physics,
Nagoya University\break
Nagoya 464-01, Japan
}
\author{
Taichiro Kugo
}
\address{
Department of Physics,
Kyoto University\break
Kyoto 606-01, Japan
}
\author{
Koichi Yamawaki
}
\address{
Department of Physics,
Nagoya University\break
Nagoya 464-01, Japan
}
\abstract{
We prove to all orders of the loop expansion
the low energy theorems of
hidden local symmetries in four-dimensional nonlinear
sigma models based on the coset space $G/H$,
with $G$ and $H$ being arbitrary compact groups.
Although the models are non-renormalizable,
the proof is done in an analogous manner to the renormalization
proof of gauge theories
and two-dimensional nonlinear sigma models
by restricting ourselves to the
operators with two derivatives
(counting a hidden gauge boson field as one derivative),
i.e., with dimension 2,
which are the only operators
relevant to the low energy limit.
Through loop-wise mathematical induction based on the
Ward-Takahashi identity for the BRS symmetry,
we solve renormalization equation for the effective action
up to dimension-2 terms plus terms with the relevant BRS sources.
We then show that all the
quantum corrections to the dimension-2 operators,
including the finite parts as well as the divergent ones,
can be entirely absorbed
into a re-definition (renormalization) of the parameters and the fields
in the dimension-2 part of the tree-level Lagrangian.
}
\endpage

\chapter{Introduction}

It is now a popular understanding that
the nonlinear
sigma model based on a coset space $G/H$ is equivalent to another
model possessing a symmetry $\Gglobal\times \Hlocal$, $\Hlocal$
being the hidden local symmetry\refmark{\BandoKugoYamawaki}.
If we further add kinetic term for the gauge
bosons of the hidden local symmetry, we obtain
phenomenological results which
are very successful in the particular case of the $\rho$ meson in the
$[SU(2)_{\rm L} \times SU(2)_{\rm R}]_{\rm global}
\times [SU(2)_{\rm V}]_{\rm local}$ model.

By choosing a parameter $a=2$ in this hidden local symmetry Lagrangian,
we have the following tree-level results for the pion and the $\rho$
meson:\refmark{\BandoKugoUeharaYamawakiYanagida}

\noindent
i) universality of the $\rho$ meson coupling\refmark{\Sakurai}
$$
g_{\rho \pi \pi} = g ~~
\hbox{($g$: hidden local gauge coupling)};
\eqno\eq
$$
ii) KSRF relation (version II)\refmark{\KSRF}
$$
m_{\rho}^2 = 2 f_{\pi}^2 g_{\rho \pi \pi}^2;
\eqno\eq
$$
iii) $\rho$ meson dominance of the electromagnetic
form factor of the pion\refmark{\Sakurai}
$$
g_{\gamma \pi \pi} = 0.
\eqno\eq
$$
{}From the tree-level Lagrangian we further obtain an $a$-independent
relation\refmark{\BandoKugoYamawakiNP}
$$
{g_{\rho}\over g_{\rho \pi \pi}} = 2 f_{\pi}^2  ,
\eqno\eq
$$
with $g_{\rho}$ being the strength of $\rho$-$\gamma$ mixing.
This coincides with another version (version I) of the celebrated
KSRF relation\refmark{\KSRF}.




The KSRF (I) is a consequence of the symmetry alone and may be
regarded as a ``low energy theorem" of the hidden local
symmetry\refmark{\BandoKugoYamawakiNP}.
Actually, the ``low energy theorem"
was proved at tree level for any Lagrangian
possessing the symmetry
and was further anticipated to survive the loop
corrections\refmark{\BandoKugoYamawakiPTP}.

Recently,
it has in fact been shown\refmark{\HaradaYamawaki}
at Landau gauge
that one-loop
effects of the pions and the $\rho$ mesons do not alter at
{\it zero momenta} the above tree-level relations of the
$[SU(2)_{\rm L} \times SU(2)_{\rm R}]_{\rm global}
\times [SU(2)_{\rm V}]_{\rm local}$ model, particularly
the "low energy theorem" mentioned above.
These results at zero momentum are
actually the relations coming from terms with
{\it two derivatives}
(counting the hidden gauge boson field as one derivative),
i.e., dimension-2 operators.
The crucial point was that the one-loop renormalization is
consistently done within those operators
{\it quite independently of the higher dimensional operators}.

Since the low energy theorem is
the statement on the {\it off-shell} amplitudes
(at zero momenta), gauge boson identification of the $\rho$ meson
in the hidden local symmetry approach is crucial:
it has a definite transformation
property and hence a definite meaning
of the off-shell extrapolation
at tree level\refmark{\BandoKugoYamawakiNP}.
However, the gauge symmetry at classical level
no longer exists at quantum
level due to gauge fixing.
Instead, there exists a Becchi-Rouet-Stora (BRS) symmetry
at quantum level as a remnant of the classical gauge symmertry.
Thus the above one-loop
results must be formulated directly as a
consequence of the BRS symmetry for the hidden local symmetry,
which in turn
yields a transparent and powerful method to analyze systematically
all orders of the loop corrections beyond one loop.

In this paper we shall prove the above low energy theorem
to all orders of the loop expansion,
based on the Ward-Takahashi (WT) identity for the
BRS symmetry.
The proof is done in the covariant gauge
and for a general case that
$G$ and $H$ are arbitrary compact groups
not restricted to the chiral case.
Particular case of the chiral group is explicitly studied
in a separate communication\refmark{\HaradaKugoYamawakiPRL}.

We are only interested in the dimension-2 operators.
Accordingly,
although the theory is
non-renormalizable and needs an infinite number of counter terms,
the proof can be done in quite the
same way as the renormalization proof
of the gauge theories\refmark{\BRS,\refZinn}
and two-dimensional nonlinear sigma
models\refmark{\BlasiCollina,\refBecchi}.
As usual we assume that there exists
a gauge invariant regularization
(for example, the dimensional regularization).

By using the loop-wise
mathematical induction based on the WT identity for the
BRS symmetry, we first derive the
renormalization equation for the $n$-th loop effective action.
Solving the
renormalization equation,
we
show that all the
quantum corrections to the dimension-2 operators can be entirely
absorbed,
{\it including the finite parts as well as the divergent ones},
into a re-definition (renormalization) of the parameters and the fields
in the dimension-2 part of the original Lagrangian.
This implies that
all the ``low energy theorems" survive loop corrections,
since they follow from the form of dimension-2 Lagrangian alone.

One might think that all the loop corrections would respect the symmetry
of the tree-level Lagrangian and thus the low energy theorem would
trivially follows and needs no ``proof".
However, the $\Hlocal$ BRS
transformation and
the $\Gglobal$ field transformation are both nonlinear
in our hidden local symmetry Lagrangian
and hence the symmetry structure of the loop corrections is far from
obvious.
In fact,
apparently non-symmetric dimension-2 operators
are induced by the loop effects.
We shall show that they are actually absorbed into a re-definition
(nonlinear point transformation) of the
Nambu-Goldstone (NG) fields.
It is the purpose of
this paper to establish that the above naive expectation,
when properly understood,
is fulfilled
even in this
highly nonlinear theory.

The paper is organized as follows.
In section 2 we present a brief review of the hidden local symmetry
for a general case that $G$ and $H$ are both arbitrary compact groups%
\refmark{\BandoKugoYamawaki,\BandoKugoYamawakiPTP}.
This is our model setting.
There we also give BRS transformations
and the precise statement of
our assertion to be proved.
Section 3 is the main body of the proof of the assertion based on the
WT identity for the BRS symmetry.
In Section 4 we show that our proof is free
from the infrared problem at least in Landau gauge
and briefly discuss the extension to the
explicit $\Gglobal$ breaking cases.
Section 5 is devoted to the summary and discussions.
Detailed steps to solve the
renormalization equation are given in Appendices A and B.


\chapter{Hidden Local Symmetry}

\section{$G/H$ algebra}

Let $G$ be a compact group
with (hermitian) generators $T_i$ satisfying
$$
[T_i,T_j] = i {f_{ij}}^k T_k , \qquad \qquad
\tr(T_iT_j) = {1\over2} \delta_{ij},
\eqn{\commutator}
$$
and $H$ be a subgroup of $G$.
Then the set of generators $\{T_i\}$ of $G$
is divided into two parts,
$\{S_a\}$ of the subgroup $H$ and
$\{X_\alpha\}$ of the rest:
$$
  \{ T_i \}
  =
  \left\{{
     S_a \in {\cal H}, \, X_\alpha \in {\cal G}-{\cal H}
  }\right\} ,
\eqn{\generators}
$$
where {\cal H} and {\cal G} denote the Lie algebra of $H$ and $G$.
It is convenient to choose the generators $\{X_\alpha\}$ of
${\cal G}-{\cal H}$ to be orthogonal to $\{S_a\}$,
$$
  \tr(S_aX_\alpha)=0 , \qquad
  \left({
    \matrix{ \tr(S_aS_b)={1\over2}\delta_{ab} \cr
             \tr(X_\alpha X_\beta)={1\over2}\delta_{\alpha\beta} \cr
    }
  }\right).
\eqn{\defoftrace}
$$
This choice implies
$\tr\left({S_a[S_b,X_\alpha]}\right) =
\tr\left({[S_a,S_b]X_\alpha}\right) = 0$
so that $[{\cal H},{\cal G}-{\cal H}] \subset {\cal G}-{\cal H}$.
Therefore the generators $\{X_\alpha\}$ of ${\cal G}-{\cal H}$
span a linear representation of $H$ which is generally reducible;
namely, $\{X_\alpha\}$ decomposes
into a set of irreducible representations
$\{X_{\alpha_k}\}(k=1,2,\ldots,n)$ such that
$$
\eqalign{
&
  h( \{X_{\alpha_1}\},\{X_{\alpha_2}\},
     \cdots,\{X_{\alpha_n}\} )h^{\dag}
\cr
& \qquad =
   ( \{X_{\beta_1}\}, \{X_{\beta_2}\}, \cdots,\{X_{\beta_n}\} )
   \left({
     \matrix{
       \rho^1_{\beta_1\alpha_1}(h) & & & 0   \cr
         & \rho^2_{\beta_2\alpha_2}(h) & &   \cr
         &    & \ddots   &                   \cr
       0 & & & \rho^n_{\beta_n\alpha_n}(h)   \cr
     }
   }\right)
}
\eqn{\reduciblerep}
$$
for ${}^{\forall} h \in H$.
Let us call the $k$-th $H$-irreducible space spanned by
$\{X_{\alpha_k}\}$, $({\cal G}-{\cal H})_k$.

We treat quite a general case for the subgroup $H$.
$H$ is generally given by a direct product
of factor groups $H_l$ each of which is simple or $U(1)$:
$$
  H = H_1 \times H_2 \times \cdots \times H_m
  \qquad
  (H_l \, : \, \hbox{\rm simple or $U(1)$} \, ) \, .
\eqn{\defofH}
$$
Thus the generators $\{S_a\}$ of ${\cal H}$
also decomposes into a set of irreducible representations
$\{S_{a_l}\}$ $(l=1,2,\cdots,m)$ of $H$;
namely, $\{S_{a_l}\}$ is a set of the generators
(adjoint representation) of the $l$-th factor group $H_l$,
which is trivial under other factor groups $H_j$ $(j\neq l)$.

\section{$\Gglobal \times \Hlocal $ model}

The nonlinear sigma model\refmark{\refCCWZ} based on
a coset $G/H$ gives an effective Lagrangian for the system
where a symmetry $G$ is spontaneously broken down
to a subgroup $H$.
Such a $G/H$ nonlinear sigma model is
generally shown\refmark{\BandoKugoYamawakiPTP}
to be gauge equivalent to another model
possessing $\Gglobal\times\Hlocal$ symmetry
in a certain limit.
The $\Gglobal \times \Hlocal $ model which we discuss
in this paper is constructed as follows%
\refmark{\BandoKugoYamawaki,\BandoKugoYamawakiPTP}.
The field variable $\xi(x)$ takes the value of
(a unitary matrix representation of) $G$,
which is parameterized as
$$
  \xi(x) = \exp [i\phi^i(x)T_i] \in G
  \quad
  (\phi^i(x) \in {\bf R})
\eqn{\defofxi}
$$
in terms of NG fields $\phi^i$
[The NG fields $\phi ^i$ split into physical ones $\phi ^\a_\perp$
corresponding to the ``broken" generators
$X_\a \in {\cal G}-{\cal H}$, and unphysical ones $\phi ^a_{\para}$
corresponding to the ``unbroken" generators
$S_a \in {\cal H}$; namely,
$\phi ^iT_i=\phi ^\a_\perp X_\a + \phi ^a_{\para}S_a$]%
\foot{
This parameterization is slightly different from
the previously used one\refmark{\BandoKugoYamawaki}\par
\noindent
$\xi(x) = \exp[i\sigma^a S_a / f_\pi]
\exp[i\pi^\alpha X_\alpha / f_\pi]$,
with $\sigma$ and $\pi$ being
the unphysical NG fields to be absorbed into
the hidden gauge bosons and the physical NG fields
living on the coset $G/H$, respectively.
}.
$\xi(x)$ transforms under $\Gglobal \times \Hlocal $ as
$$
  \xi(x) \rightarrow
  \xi'(x) = h(x) \xi(x) g^{\dag} ,
  \quad
  g \in \Gglobal ,
  \quad
  h(x) \in \Hlocal .
\eqn{\defoftrans}
$$
A basic quantity is a (covariantized) Maurer-Cartan 1-form:
$$
  \hat{\alpha}_\mu(x) =
  D_\mu \xi(x) \cdot \xi^{\dag}(x) / i ,
\eqn{\MaurerCartan}
$$
where $D_\mu$ is $H$-covariant derivative given by
$$
  D_\mu \xi(x) = \partial_\mu \xi(x) - i V_\mu(x) \xi(x) ,
\eqn{\covariatderivative}
$$
with $V_\mu \equiv V_\mu^aS_a$ being the hidden gauge boson.
The Maurer-Cartan 1-form $\hat{\alpha}_\mu$ is Lie-algebra valued
($\in {\cal G}$) and transforms under the
$\Gglobal \times \Hlocal $ transformation \defoftrans \ as
$$
  \hat{\alpha}_\mu(x)
  \rightarrow
  \hat{\alpha}'_\mu(x)
  =
  h(x) \hat{\alpha}_\mu(x) h^{\dag}(x);
\eqn{\transofMaurer}
$$
namely, it is $\Gglobal $-invariant and transforms homogeneously
under $\Hlocal $.
Therefore it splits into two parts,
$\hat{\alpha}_{\mu\parallel}$ belonging to ${\cal H}$ and
$\hat{\alpha}_{\mu\perp}$ belonging to ${\cal G}-{\cal H}$,
and each of them is further decomposed into the above-mentioned
$H$-irreducible pieces,
$\hat{\alpha}_{\mu\parallel}^{(l)}\,(l=1,\cdots,m)$ and
$\hat{\alpha}_{\mu\perp}^{(k)}\,(k=1,\cdots,n)$:
$$
\eqalign{
  \hat{\alpha}_{\mu\parallel}^{(l)}(x)
&\equiv
  \sum_{a_l\in{\cal H}_l}
  2 S_{a_l}
  \tr\left({
    S_{a_l}\hat{\alpha}_\mu(x)
  }\right) ,
\cr
  \hat{\alpha}_{\mu\perp}^{(k)}(x)
&\equiv
  \sum_{\alpha_k\in({\cal G}-{\cal H})_k}
  2 X_{\alpha_k}
  \tr\left({
    X_{\alpha_k}\hat{\alpha}_\mu(x)
  }\right) .
}
\eqno\eq
$$
Thus the most general $\Gglobal \times \Hlocal $ invariant
Lagrangian which contains the least number of derivatives
is given by
$$
\eqalignno{
  {\cal L}=
&
  \sum^m_{l=1}a_{\parallel l} {\cal L}_{\parallel}^{(l)}
  + \sum^n_{k=1}a_{\perp k} {\cal L}_{\perp}^{(k)}
  + {\cal L}_{\rm kin}(V_\mu) ,
&\eqname{\HiddenLagrangian}
\cr
&
  {\cal L}_{\parallel}^{(l)}
  \equiv
  f_\pi^2 \tr
  \left[{
    \left({
      \hat{\alpha}_{\mu\parallel}^{(l)}
    }\right)^2
  }\right] ,
&\eq
\cr
&
  {\cal L}_{\perp}^{(k)}
  \equiv
  f_\pi^2 \tr
  \left[{
    \left({
      \hat{\alpha}_{\mu\perp}^{(k)}
    }\right)^2
  }\right] ,
&\eq
\cr
&
  {\cal L}_{\rm kin}(V_\mu)
  \equiv
  \sum_{l=1}^m - {1\over4g_l^2}
  F_{\mu\nu}^2(V_\mu^{(l)}) ,
&\eq
}
$$
where $a_{\parallel l}$, $a_{\perp k}$, $g_l$ being arbitrary
parameters, and $F_{\mu\nu}(V_\mu^{(l)})$ is the field strength
of the hidden gauge field $V_\mu^{(l)}$
of the $l$-th factor group $H_l$.
This is the hidden local symmetry Lagrangian.

In many applications, we often need to couple the system to some
external gauge bosons by gauging a part of the $\Gglobal $ group.
For instance, the electromagnetic interaction is introduced by
gauging the $U(1)_{\rm em}$ part of the chiral group
$[SU(2)_{\rm L}\times SU(2)_{\rm R}]_{\rm global}$.
For the present purpose to discuss generically
the renormalization effects
in the nonlinear hidden local symmetry Lagrangian,
it is convenient to gauge the full $\Gglobal $ group.
So we introduce the external gauge field
${\cal V}_\mu(x)\equiv{\cal V}_\mu^i(x)T_i$ of
$\Gglobal $ (which is now a local group despite the name),
and replace the above covariant derivative \covariatderivative \ by
$$
  D_\mu \xi(x) =
  \partial_\mu \xi(x) - i V_\mu(x) \xi(x)
  + i \xi(x) {\cal V}_\mu .
\eqno\eq
$$
Then the hidden local symmetry Lagrangian \HiddenLagrangian \ is
invariant under
both local $\Gglobal \times \Hlocal $ transformation:
$$
  \xi(x) \rightarrow \xi'(x)
  =
  h(x) \xi(x) g^{\dag}(x) .
\eqn{\FullTrans}
$$

For later reference, we note that,
in the Lagrangian \HiddenLagrangian,
the hidden gauge field $V_\mu^{(l)}=\sum_{a_l}V_\mu^{a_l}S_{a_l}$
of the $l$-th factor group $H_l$ appears only
in ${\cal L}_{\parallel}^{(l)}$
(aside from ${\cal L}_{\rm kin}(V_\mu)$)
in the form:
$$
\eqalignno{
  {\cal L}_\parallel^{(l)}
&=
  f_\pi^2 {1\over2} \sum_{a_l\in {\cal H}_l}
  (V_\mu^a - \Apara_\mu^a )^2 ,
&\eqname{\Lparallel}
\cr
  \Apara_\mu^a
&=
  {2\over i} \tr
  \left[{
    S^a (\partial_\mu\xi + i \xi {\cal V}_\mu ) \xi^{\dag}
  }\right] .
&\eqname{\Amupart}
}
$$

In the absence of the gauge-field kinetic term
${\cal L}_{\rm kin}(V_\mu)$,
the hidden local symmetry Lagrangian
\HiddenLagrangian\ is equivalent to the
usual nonlinear sigma model based on $G/H$;
indeed, then ${\cal L}_{\parallel}^{(l)}$
in \Lparallel\ vanishes by use of
the $V_\mu^a$ equation of motion,
$V_\mu^a = \Apara_\mu^a $, and \HiddenLagrangian\
reduces to the $G/H$ Lagrangian
$\sum^n_{k=1}a_{\perp k} {\cal L}_{\perp}^{(k)}$%
\refmark{\refCCWZ,\refBB}
[Note that we are using $f_\pi$ as
a unique dimensionful constant and
the usual ``decay constants'' of the
true NG fields $\phi_{\perp}^i$ are given by
$\left(a_{\perp k}\right)^{1/2}f_\pi$.].

\section{BRS transformation}

In this paper we do not consider the radiative corrections
due to the external gauge field ${\cal V}_\mu$
regarding its coupling as weak.
So we do not add the kinetic term nor the gauge-fixing term
for ${\cal V}_\mu$.
As for the gauge-fixing for the hidden gauge boson $V_\mu$,
we take the covariant gauge, so that the gauge-fixing and
Faddeev-Popov (FP) terms are given by
$$
  {\cal L}_{\rm GF} + {\cal L}_{\rm FP}
  =
  B^a \partial^\mu V_\mu^a
  + {1\over2} \alpha_l g_l^2 B^a B^a
  + i \bar{C}^a \partial^\mu D_\mu C^a \ ,
\eqn{\GFFPlagrangian}
$$
where $\alpha_l$ is a gauge parameter for the factor group
$H_l$ and we have used a shorthand notation
$$
\alpha_l g_l^2 B^a B^a \equiv
\sum_l \alpha_l g_l^2 \sum_{a_l\in{\cal H}_l} B^{a_l}B^{a_l}\ .
\eqno\eq
$$

The infinitesimal form of the $\Gglobal \times \Hlocal $
transformation \FullTrans $\,$ is given by
$$
\eqalignno{
  \delta \xi(x)
&=
  i \theta(x) \xi(x) - i \xi(x) \vartheta(x) ,
&\eqname{\InfTransofXi}
\cr
&  \theta(x)\equiv \theta^a(x)S_a, \quad
  \vartheta(x)\equiv \vartheta^i(x)T_i.
& \
}
$$
This defines the transformation of the field $\phi^i(x)$ in \defofxi,
$\xi(x) = \exp [i\phi^i(x)T_i]$, in the form
$$
\eqalign{
  \delta \phi^i(x)
&=
  \theta^a W_a^i(\phi) + \vartheta^j {\cal W}_j^i(\phi)
\cr
&\equiv
  \bftheta^A {\mib W}_A^i(\phi)
  \equiv
  \bftheta^A \hat{\mib W}_A \phi^i
  \qquad
  (\hat{\mib W}_A \equiv {\mib W}_A^i(\phi)
  {\partial\over\partial\phi^i} ),
}
\eqno\eq
$$
where $A=(a,i)$ denotes a set of labels of the $\Hlocal $
and $\Gglobal $ generators;
$\bftheta^A\equiv(\theta^a,\vartheta^i)$
and $\hat{\mib W}_A\equiv(\hat{W}_a,\hat{{\cal W}}_i)$.
By this definition, these generators $\hat{\mib W}_A$
clearly satisfy the algebra of
$\Hlocal \times \Gglobal $:
$$
\eqalignno{
&
  [\hat{\mib W}_A,\hat{\mib W}_B] = {f_{AB}}^C \hat{\mib W}_C \, ,
&\eqname{\RelOfAlgebra}
\cr
&\quad \hbox{\rm i.e.,} \quad
  \left\{{
    \matrix{
      [\hat{W}_a,\hat{W}_b] = {f_{ab}}^c \hat{W}_c ,\hfill\, \cr
      [\hat{{\cal W}}_i,\hat{{\cal W}}_j]
        = {f_{ij}}^k \hat{{\cal W}}_k , \hfill\, \cr
      [\hat{W}_a,\hat{{\cal W}}_i] = 0 . \hfill\,
    }
  }\right.
& \
}
$$
It is important that we have the same number of $\Gglobal $
transformation generators $\hat{{\cal W}}_i$ as our field
variables $\phi^i$ and $\hat{{\cal W}}_i$ take the form
$$
  \hat{{\cal W}}_i
  =
  - {\partial \over \partial \phi^i}
  + {\cal O}(\phi) \times
  {\partial \over \partial \phi} .
\eqn{\DefcalW}
$$
Indeed, as is easily shown we have
$$
  \hat{{\cal W}}_i F(\phi) = 0
  \ \hbox{for}\, {}^\forall i
  \quad
  \Rightarrow
  \quad
  F(\phi) = \hbox{\rm const. ($\phi$-independent)} .
\eqn{\Solzero}
$$
Another point to be noted here is that if we set
$\vartheta=\theta$,
(i.e., $\vartheta^a=\theta^a$ and $\vartheta^\alpha=0$ when writing
$\vartheta = \vartheta^aS_a + \vartheta^\alpha X_\alpha$ ),
the transformation \InfTransofXi \ becomes
a linear $H$-transformation on $\phi$; therefore we have
$$
  \left({
    \hat{W}_a + \hat{{\cal W}}_a
  }\right)
  \phi^i
  =
  \phi^j {f_{ja}}^i \, .
\eqn{\Hdiagtrans}
$$
We call this linear transformation $\Hdiag $-transformation,
and the covariance under $\Hdiag $ provides us
with a useful tool below.

The BRS transformation is given simply by replacing the infinitesimal
transformation parameter $\bftheta^A=(\theta^a,\vartheta^i)$
by the FP ghost field ${\mib C}^A=(C^a,{\cal C}^i)$:
$$
  \dBRS \phi^i
  =
  \left({
    C^a \hat{W}_a + {\cal C}^j \hat{{\cal W}}_j
  }\right)
  \phi^i
  =
  {\mib C}^A \hat{\mib W}_A \phi^i ,
\eqn{\PhitransB}
$$
where ${\cal C}^i$ is the FP ghost for the external $\Gglobal $ gauge
field ${\cal V}_\mu^i$.
Note that ${\cal C}^i$ is a non-propagating field, since we are not
quantizing ${\cal V}_\mu^i$.
The nilpotency requirement ${(\dBRS)}^2=0$ on $\phi^i$
with the algebra \RelOfAlgebra \ determines
the FP ghost BRS transformation as usual:
$$
  \dBRS {\mib C}^A
  =
  -{1\over2}
  {\mib C}^B {\mib C}^C {f_{BC}}^A .
\eqn{\CtransB}
$$
The BRS transformation of the $\Hlocal\times\Gglobal$
gauge fields $\VA_\mu^A \equiv (V_\mu^a,{\cal V}_\mu^i)$
is of course given by
$$
  \dBRS \VA_\mu^A
  =
  \partial_\mu {\mib C}^A + \VA_\mu^B {\mib C}^C {f_{BC}}^A .
\eqno\eq
$$

\section{Assertion}

For definiteness, let us first define the dimensions
of our field as follows:
$$
  \dim [\phi^i]=0 , \quad
  \dim[V_\mu^a]=\dim[{\cal V}_\mu^i]=1 .
\eqn{\DimofField}
$$
These are canonical dimensions,
since we are using the parameterization
$\xi=\exp(i\phi^iT_i)$, and the gauge fields appear in the covariant
derivative $D_\mu$ (of dimension 1).
It is also convenient to assign the following dimensions
to the FP-ghosts:
$$
  \dim[C^a]=\dim[{\cal C}^i]=0, \quad
  \dim[\bar{C}_a]=2.
\eqn{\DimofGhost}
$$
Then the BRS transformation preserves the dimensions.

In this terminology,
our hidden local symmetry Lagrangian
\HiddenLagrangian\  with \GFFPlagrangian\
consists of two parts, dimension-2 part
$\sum_l a_{\parallel l} {\cal L}_{\parallel}^{(l)} +
\sum_k a_{\perp k} {\cal L}_{\perp}^{(k)}$ and
dimension-4 part
${\cal L}_{\rm kin}(V_\mu) + {\cal L}_{\rm GF} + {\cal L}_{\rm FP}$.
[Here we mean the field plus derivative dimensions.]
We consider the quantum corrections to this system
at any loop order.
What we wish to prove in this paper is the following proposition:

{\bf Proposition}
{\it As far as the dimension-2 operators are concerned,
all the quantum corrections,
including the finite parts as well as the divergent parts,
can be absorbed into the original dimension-2 Lagrangian
$\sum_l a_{\parallel l} {\cal L}_{\parallel}^{(l)} +
\sum_k a_{\perp k} {\cal L}_{\perp}^{(k)}$
by a suitable redefinition (renormalization) of the parameters
$a_{\parallel l}$, $a_{\perp k}$, and the fields $\phi^i$,
$V_\mu^a$.}

Namely, this implies that the tree-level dimension-2 Lagrangian,
with the parameters and fields substituted by the
``renormalized" one, already
describes the exact action at any loop order,
and therefore that all the ``low-energy theorems'' derived from it
receive no quantum corrections at all.

\chapter{Proof of the Proposition}

\section{Ward-Takahashi identity}

The proof of our proposition goes in quite the same way
as the renormalizability proofs for the
gauge theories\refmark{\BRS,\refZinn}
and the nonlinear Lagrangians\refmark{\BlasiCollina,\refBecchi}.
Actually, our hidden local symmetry
Lagrangian is a combined system
of the gauge theory and the nonlinear Lagrangian,
and hence the proof can be done
by use of the techniques for both of them.

Following the usual procedure,
we can write down the Ward-Takahashi (WT) identity
for the effective action $\Gamma$
both for the gauged-$\Gglobal$ and $\Hlocal$ symmetries.
We also make a usual assumption that
there exists a gauge invariant regularization.
The Nakanishi-Lautrap fields $B^a$
and the FP anti-ghost fields $\bar{C}^a$ can be eliminated
from $\Gamma$ through their equations of motion as usual.
After eliminating them,
the tree level action $S=\Gamma_{\rm tree}$ reads
$$
\eqalignno{
&
  S [ \Phi, \Kex; {\mib a} ]
  =
  S_2 [\phi,\VA] + S_4[\Phi,\Kex] ,
&\eqname{\TreeAction}
\cr
& \qquad
  S_2 [\phi,\VA] = \int d^4x
  \left({
    \sum_l a_{\parallel l} {\cal L}_{\parallel}^{(l)} (\phi,\VA)
    + \sum_k a_{\perp k} {\cal L}_{\perp}^{(k)} (\phi,\VA)
  }\right) ,
& \
\cr
& \qquad
  S_4 [\Phi,\Kex] = \int d^4x
  \left({
    {\cal L}_{\rm kin}(V_\mu) + K_i \dBRS \phi^i
    + {\mib K}_A^\mu \dBRS \VA_\mu^A + {\mib L}_A \dBRS {\mib C}^A
  }\right) ,
&\eqname{\SfourAction}
}
$$
with collective notations
${\mib a} \equiv (a_{\parallel l},a_{\perp k})$,
$\VA_\mu^A \equiv (V_\mu^a,{\cal V}_\mu^i)$,
${\mib K}_A^\mu \equiv (K_a^\mu,{\cal K}_i^\mu)$,
${\mib L}_A \equiv (L_a,{\cal L}_i)$,
$\Phi \equiv (\phi^i,\VA_\mu^A,{\mib C}^A)$ and
$\Kex \equiv (K_i,{\mib K}_A^\mu,{\mib L}_A)$.
Note that according to the field dimension assignment
\DimofField \ and \DimofGhost,
the dimension of the BRS source $\Kex$ is given by
$$
  \dim[K_i]=\dim[{\mib L}_A]=4, \quad
  \dim[{\mib K}_A^\mu]=3 .
\eqno\eq
$$
Then $S_2$ and $S_4$ in the action \TreeAction \
stand for the parts carrying the (field plus derivative)
dimension 2 and 4, respectively.
The WT identity for the effective action $\Gamma$ is given by
$$
  \Gamma \ast \Gamma = 0 ,
\eqn{\WTidentity}
$$
with the $\ast$ operation defined by
$$
  F \ast G
  =
  {(-)}^\Phi
  {\overleftarrow{\delta} F \over \delta \Phi}
  {\delta G \over \delta \Kex}
  - {(-)}^\Phi
  {\overleftarrow{\delta} F \over \delta \Kex}
  {\delta G \over \delta \Phi}
\eqno\eq
$$
for arbitrary functionals $F[\Phi,\Kex]$ and $G[\Phi,\Kex]$.
(Here the symbols $\delta$ and $\overleftarrow{\delta}$
denote the derivatives from the left and right, respectively,
and ${(-)}^\Phi$ denotes $+1$ or $-1$
when $\Phi$ is bosonic or fermionic, respectively.)

The effective action $\Gamma$ is calculated in the loop expansion:
$$
  \Gamma = S + \hbar \Gamma^{(1)} + \hbar^2 \Gamma^{(2)} + \cdots .
\eqno\eq
$$
Actually, to calculate the renormalized $\Gamma$
to $n$-th loop order, we need to use the following ``bare" action
${(S_0)}_n$ which is obtained by substituting
the $n$-th loop order ``bare" fields ${(\Phi_0)}_n$, ${(\Kex_0)}_n$
and parameters ${({\mib a}_0)}_n$ into the tree-level action
$S[\Phi,\Kex;{\mib a}]$ in \TreeAction:
$$
  {(S_0)}_n = S
  \left[{
    {(\Phi_0)}_n , {(\Kex_0)}_n ; {({\mib a}_0)}_n
  }\right] ,
\eqno\eq
$$
$$
\eqalign{
  {(\Phi_0)}_n
&=
  \Phi + \hbar \delta \Phi^{(1)} + \cdots +
  \hbar^n \delta \Phi^{(n)} ,
\cr
  {(\Kex_0)}_n
&=
  \Kex + \hbar \delta \Kex^{(1)}
  + \cdots + \hbar^n \delta \Kex^{(n)} ,
\cr
  {({\mib a}_0)}_n
&=
  {\mib a} + \hbar \delta{\mib a}^{(1)} +
  \cdots + \hbar^n \delta{\mib a}^{(n)}
}
\eqn{\FieldRenorm}
$$
[Note that the renormalization of the decay constants
$\left(a_{\parallel l}\right)^{1/2}\!\!f_\pi$,
$\left(a_{\perp k}\right)^{1/2}\!\!f_\pi$ is
performed on the parameter ${\mib a}$,
so that our unique dimensionful parameter $f_\pi$ is
kept fixed.].
Therefore, the WT identity \WTidentity\  for $\Gamma$ calculated
based on this ``bare" action in fact should read
$$
  {(-)}^\Phi
  { \overleftarrow{\delta} \Gamma \over \delta {(\Phi_0)}_n }
  { \delta \Gamma \over \delta {(\Kex_0)}_n }
  - {(-)}^\Phi
  { \overleftarrow{\delta} \Gamma \over \delta {(\Kex_0)}_n }
  { \delta \Gamma \over \delta {(\Phi_0)}_n }
  = 0 .
\eqno\eq
$$
However we shall show below
that the field renormalization \FieldRenorm,
$(\Phi,\Kex) \rightarrow
\left({ {(\Phi_0)}_n, {(\Kex_0)}_n }\right)$,
is a ``canonical" transformation for any $n$
such that the ``Poisson bracket" $F \ast G$ remains invariant:
$$
  {(-)}^\Phi
  { \overleftarrow{\delta} F \over \delta \Phi }
  { \delta G \over \delta \Kex }
  - {(-)}^\Phi
  { \overleftarrow{\delta} F \over \delta \Kex }
  { \delta G \over \delta \Phi }
  =
  {(-)}^\Phi
  { \overleftarrow{\delta} F \over \delta {(\Phi_0)}_n }
  { \delta G \over \delta {(\Kex_0)}_n }
  - {(-)}^\Phi
  { \overleftarrow{\delta} F \over \delta {(\Kex_0)}_n }
  { \delta G \over \delta {(\Phi_0)}_n } .
\eqno\eq
$$
So we always have $\Gamma\ast\Gamma=0$ written in terms of
the tree level fields $\Phi$ and $\Kex$.

Now the ${\cal O}(\hbar^n)$ term $\Gamma^{(n)}$,
which contains not only the genuine $n$-loop terms but also
the contributions of lower loop diagrams with counter terms,
is further expanded according to the dimensions:
$$
  \Gamma^{(n)} =
  \Gamma_0^{(n)} [\phi] + \Gamma_2^{(n)}[\phi,\VA]
  + \Gamma_4^{(n)}[\Phi,\Kex] + \Gamma_6^{(n)}[\Phi,\Kex]
  + \cdots .
\eqn{\NthAction}
$$
Here again we are counting the dimensions of only the fields
and derivatives. [The deviation from dimension-4 is
compensated by powers of the unique dimensionful parameter $f_\pi^2$.]
The first dimension-0 term $\Gamma_0^{(n)}$ can contain only
the dimensionless field $\phi^i$ without derivatives.
The dimension-2 of the second term $\Gamma_2^{(n)}$ is
supplied by derivative and/or the gauge field $\VA_\mu^A$.
The BRS source field $\Kex$, carrying dimension 4 or 3,
can appear only in $\Gamma_4^{(n)}$ and beyond:
the dimension-4 term $\Gamma_4^{(n)}$ is at most linear in $\Kex$,
while the dimension-6 term $\Gamma_6^{(n)}$ can
contain not only  linear terms in $\Kex$ but
also a quadratic term in $K_a^\mu$,
the BRS source of the hidden gauge boson $V_\mu^a$.

\section{Proof of the Assertion}

Let us now prove the following by mathematical induction with respect
to the loop expansion order $n$: for $n=1,2,\cdots,$

\item{1)} $\Gamma_0^{(n)} = 0$.

\item{2)}
By choosing suitably the $n$-th order counter terms
$\delta\Phi^{(n)}$, $\delta\Kex^{(n)}$ and $\delta{\mib a}^{(n)}$
in \FieldRenorm,
$\Gamma_2^{(n)}[\phi,\VA]$ and the $\Kex$-linear terms
in $\Gamma_4^{(n)}[\Phi,\Kex]$
can be made vanish;
$\Gamma_2^{(n)} [\phi,\VA] =
\Gamma_4^{(n)}[\Phi,\Kex]\vert_{\Klinear} =0$.

\item{3)}
The field reparameterization (renormalization)
$(\Phi,\Kex)\rightarrow
\left({ {(\Phi_0)}_n,{(\Kex_0)}_n }\right)$ is a ``canonical"
transformation which leaves the $\ast$ operation invariant.

Suppose that these statements hold up to $n-1$,
and calculate the $n$-th loop effective action
$\Gamma^{(n)}$, for the moment,
using the $(n-1)$-st loop level ``bare" action
${(S_0)}_{n-1}$ [namely, without $n$-th loop level counter terms
$\delta\Phi^{(n)}$, $\delta\Kex^{(n)}$ and $\delta{\mib a}^{(n)}$].
Then we have the WT identity \WTidentity\ thanks to the induction
assumption 3), which yields for $\hbar^n$ terms
$$
  S \ast \Gamma^{(n)} =
  - {1\over2}
  \sum_{l=1}^{n-1} \Gamma^{(l)} \ast \Gamma^{(n-l)} .
\eqn{\NthWTidentity}
$$
Substituting the dimensional expansions,
$S=S_2+S_4$ [\TreeAction] for $S$ and \NthAction\
for $\Gamma^{(l)}$ $(l=1,\cdots,n)$,
we compare both sides of \NthWTidentity\
possessing the same dimension.
Since $\Gamma_0^{(l)}$ and $\Gamma_2^{(l)}$ vanish for
$1 \leq l \leq n-1$ by the induction assumption,
there are no dimension 0 and 2 parts
in the RHS of \NthWTidentity, so that we have
$$
\eqalignno{
  \dim0 \, : \quad
&
  \quad
  S_4 \ast \Gamma_0^{(n)} + S_2 \ast \Gamma_2^{(n)} = 0 ,
&\eqname{\DimzeroWT}
\cr
  \dim2 \, : \quad
&
  \quad
  S_4 \ast \Gamma_2^{(n)} + S_2 \ast \Gamma_4^{(n)} = 0 .
&\eqname{\DimtwoWT}
}
$$
[Note that the $\ast$ operation lowers the dimension by 4.]
For dimension 4 parts, the RHS in \NthWTidentity\  might seem to
have contribution of the form
$\sum_{l=1}^{n-1} \Gamma_4^{(l)} \ast \Gamma_4^{(n-l)}$,
but they actually vanish
since all the $\Gamma_4^{(l)}$ $(1 \leq l \leq n-1)$ contain
no $\Kex$ again by the induction assumption.
So, we find
$$
  \dim4 \, : \quad \quad
  S_4 \ast \Gamma_4^{(n)} + S_2 \ast \Gamma_6^{(n)} = 0 .
\eqn{\DimfourWT}
$$
These three equations (renormalization equations)
\DimzeroWT\ -- \DimfourWT\
give enough information
for determining the possible forms of $\Gamma_0^{(n)}$,
$\Gamma_2^{(n)}$ and
$\Gamma_4^{(n)}\vert_{\Klinear}$
(the $\Kex$-linear term in $\Gamma_4^{(n)}$)
which we are interested in.

Noting that the BRS transformation $\dBRS$ on the fields
$\Phi=(\phi^i,\VA_\mu^A,{\mib C}^A)$ can be written in the form
$$
  \dBRS =
  {\delta S_4 \over \delta \Kex }
  {\delta \over \delta \Phi} ,
\eqno\eq
$$
we see it convenient to define an analogous transformation
$\delta'_\Gamma$ on the fields $\Phi$ by
$$
  \delta'_\Gamma =
  {\delta \Gamma^{(n)}_4 \over \delta \Kex }
  {\delta \over \delta \Phi} .
\eqno\eq
$$
Then we can write $\Gamma_4^{(n)}$ in the form
$$
  \Gamma_4^{(n)} =
  A_4[\phi,\VA] + K_i \delta'_\Gamma \phi^i
  + {\mib K}_A^\mu \delta'_\Gamma \VA_\mu^A +
  {\mib L}_A \delta'_\Gamma {\mib C}^A .
\eqno\eq
$$
In terms of this notation, \DimzeroWT\ -- \DimfourWT\
can be rewritten into
$$
\eqalignno{
  & \dBRS \Gamma_0^{(n)} = 0 ,
&\eqname{\DimzeroWTB} \cr
  & \dBRS \Gamma_2^{(n)} + \delta'_\Gamma S_2 = 0 ,
&\eqname{\DimtwoWTB} \cr
  & \dBRS \Gamma_4^{(n)} + \delta'_\Gamma S_4
  + {\delta \Gamma_6^{(n)} \over \delta \Kex}
    {\delta S_2 \over \delta \Phi} = 0 ,
&\eqname{\DimfourWTB}
}
$$
respectively,
where use has been made of the fact that $S_2$,
$\Gamma_0^{(n)}$ and $\Gamma_2^{(n)}$ contain no $\Kex$.

{}From \DimzeroWTB \ it immediately follows that $\Gamma_0^{(n)}=0$.
This is because $\Gamma_0^{(n)}[\phi]$ is a function of only $\phi^i$
containing no derivatives and the BRS transformation $\dBRS$
on such a function is just a
$\Gglobal  \times \Hlocal $ transformation,
but we know that there is no $\Gglobal  \times \Hlocal $-invariant
without derivatives.
Thus our first statement
1) $\Gamma_0^{(n)}=0$, in the above, has been proved.

To prove the second and the third statements  2) and 3),
we need to solve the equations \DimtwoWTB \ and \DimfourWTB,
which is much more non-trivial task.
A bit lengthy analysis of \DimtwoWTB \ and \DimfourWTB,
which is given in Appendices,
shows that the general solution is given in the form
$$
  \Gamma_2^{(n)} +
  \left. \Gamma_4^{(n)} \right\vert_{\Klinear}
  =
  A_{2\GI}^{(n)} [\phi,\VA] - S \ast Y^{(n)}
\eqn{\Solution}
$$
up to irrelevant terms
(dimension-6 or $\Kex$-independent dimension-4 terms).
Here $A_{2\GI}$ is a dimension-2 gauge-invariant function
of $\phi^i$ and $\VA_\mu^A$ and
$$
  Y^{(n)} =
  \int d^4x
  \left[{
    K_i F^{(n)i} (\phi)
    + \alpha_l^{(n)} K_a^\mu V_\mu^a
    + \beta_l^{(n)} L_a C^a
    + \gamma_l^{(n)} f_{abc} K_a^\mu K_{b\mu} C^c
  }\right] ,
\eqn{\SolY}
$$
where $F^{(n)i}$ are arbitrary functions
and $\alpha_l^{(n)}$, $\beta_l^{(n)}$ and $\gamma_l^{(n)}$
are arbitrary constants.
In \SolY\  we have used shorthand notations like
$$
  \alpha_l^{(n)} K_a^\mu V_\mu^a
  \equiv
  \sum_{l=1}^m \alpha_l^{(n)}
  \left({
    \sum_{a\in {\cal H}_l} K_a^\mu V_\mu^a
  }\right) ;
\eqno\eq
$$
namely, the parameters $\alpha_l^{(n)}$, $\beta_l^{(n)}$
and $\gamma_l^{(n)}$ in \SolY \ can take different values
for different factor groups $H_l$ in
$H=H_1 \times H_2 \times \cdots H_m$.

The form \Solution\ of the solution already proves
our desired statements 2) and 3) in the above, as seen as follows.
First recall that the above $\Gamma^{(n)}$ was calculated using
$\left({S_0}\right)_{n-1}$ without $n$-th loop level
counter terms $\delta\Phi^{(n)}$, $\delta\Kex^{(n)}$
and $\delta{\mib a}^{(n)}$.
If we include those, we have the following additional contributions
to $\Gamma^{(n)}$:
$$
  \Delta \Gamma^{(n)}
  =
  \delta\Phi^{(n)} {\delta S \over \delta \Phi}
  + \delta\Kex^{(n)} {\delta S \over \delta \Kex}
  + \delta{\mib a}^{(n)} {\partial S \over \partial {\mib a}} \, ,
\eqn{\AddCont}
$$
with $S[\Phi,\Kex;{\mib a}]$ being the tree-level action \TreeAction.
So the true $n$-th loop level effective action is
$\Gamma^{(n)}+\Delta\Gamma^{(n)}\equiv\Gamma_{\rm total}^{(n)}$.
We now show that these $n$-th loop level counter terms
can be chosen such that $\Gamma_{\rm total}^{(n)}$ has
no dimension-2 and dimension-4-$\Kex$-linear terms;
namely, the quantities in \Solution \ can be
completely canceled by $\Delta\Gamma^{(n)}$.

First is the $A_{2\GI}^{(n)}[\phi,\VA]$ term.
Since we know that ${\cal L}_{\parallel}^{(l)}$
and ${\cal L}_{\perp}^{(k)}$ span a complete set
of dimension-2 $\Gglobal  \times \Hlocal $ gauge invariants,
$A_{2\GI}^{(n)}$ must be a linear combination of them:
$$
  A_{2\GI}^{(n)}
  =
  \int d^4x
  \left({
    \sum_l b_{\parallel l}^{(n)} {\cal L}_{\parallel}^{(l)} (\phi,\VA)
    + \sum_k b_{\perp k}^{(n)} {\cal L}_{\perp}^{(k)} (\phi,\VA)
  }\right) ,
\eqno\eq
$$
with ${\mib b}^{(n)}\equiv(b_{\parallel l}^{(n)},b_{\perp k}^{(n)})$
being certain coefficients.
But this can be written as
${\mib b}^{(n)} \cdot \partial S/\partial {\mib a}$ and can just be
canceled by choosing the $a$-parameter counter terms
$\delta{\mib a}^{(n)}=
(\delta a_{\parallel l}^{(n)},\delta a_{\perp k}^{(n)})$
as $\delta{\mib a}^{(n)}=-{\mib b}^{(n)}$.

Next consider the $-S\ast Y^{(n)}$ term.
[This term includes
the gauge non-invariant dimension-2 operators
$$
  -
  \left({
    F^{(n)i}(\phi)
    {\delta \over \delta \phi^i} S_2
     + \alpha_l^{(n)} V_\mu^a
    {\delta \over \delta V_\mu^a} S_2
  }\right)
\eqno\eq
$$
generated by the loop effects\foot{
It was pointed out that
in the nonlinear sigma model (without hidden gauge bosons)
such non-invariant terms are generated by the one-loop effects
for dimension-4 operators,
which are transformed away by the NG-field redefinition involving
space-time derivatives\refmark{\refAppBer}.
In this case without propagating gauge bosons
there exist no loop corrections
to the dimension-2 operators
(up to quadratic divergences which are absent
in the dimensional regularization).
}
(see (A.58)).]
We should note that this term $-S\ast Y^{(n)}$ just represents
a ``canonical transformation" of $S$ caused by $-Y^{(n)}$
as its generating functional.
It is therefore clear that
if we choose the $n$-th order field counter terms
$\delta\Phi^{(n)}$ and $\delta\Kex^{(n)}$
to be equal to the canonical transformations of $\Phi$ and $\Kex$
generated by $+Y^{(n)}$,
$$
\eqalign{
  \delta\Phi^{(n)}
&=
  \Phi \ast Y^{(n)} =
  {(-)}^\Phi{\delta Y^{(n)} \over \delta \Kex} ,
\cr
  \delta\Kex^{(n)}
&=
  \Kex\ast Y^{(n)} =
  -{(-)}^\Phi {\delta Y^{(n)} \over \delta \Phi} ,
}
\eqn{\RenormField}
$$
then the additional contributions in \AddCont \ yield
$$
  \delta\Phi^{(n)} {\delta S \over \delta \Phi}
  + \delta\Kex^{(n)} {\delta S \over \delta \Kex}
  = S \ast Y^{(n)}
\eqno\eq
$$
and cancels the $-S\ast Y^{(n)}$ term.
Eq.\RenormField \ also shows that the field counter terms
$\delta\Phi^{(l)}$ and $\delta\Kex^{(l)}$
($l=1,2,\cdots,n-1$) at lower loop levels,
which are determined at the preceding steps of this induction argument,
are also generated by certain generating functional $Y^{(l)}$.
Thus the field transformation
$(\Phi,\Kex)\rightarrow
\left({ {(\Phi_0)}_n , {(\Kex_0)}_n }\right)$
is an infinitesimal ``canonical transformation" generated by
$\sum_{l=1}^n Y^{(l)}$,
so that the ``Poisson bracket" $F \ast G$ remains invariant.
This completes the proof of our statements 1) -- 3),
and hence our Proposition.

At this point let us comment
on the nature of the field renormalization
\RenormField.
The renormalization of the hidden gauge boson $V_\mu$
is given by
$ \delta V_\mu^a =
  \delta Y^{(n)} / \delta K_a^\mu
  = \alpha^{(n)}_l V_\mu^a$.
Thus $V_\mu$ is
{\it multiplicatively renormalized}%
\foot{
There is in fact an ambiguity in the expression \Solution;
a gauge-invariant term
$-2\epsilon
a_{\parallel l} {\cal L}_{\parallel}^{(l)}$
in $A_{2\GI}^{(n)}$ can also be
written in the form
$S\ast[\epsilon K_a^\mu (V_\mu^a - \Apara_\mu^a)]$.
Then the $K_a^\mu$ term in $Y^{(n)}$
,\SolY,
is replaced by
$K_a^\mu [\alpha_l^{(n)} V_\mu^a +
\epsilon (V_\mu^a - \Apara_\mu^a)]$.
This form would imply a mixing of
$V_\mu^a - \Apara_\mu^a$ with $V_\mu^a$
through the renormalization;
$\delta V_\mu^a = \alpha_l^{(n)} V_\mu^a +
\epsilon (V_\mu^a - \Apara_\mu^a)$,
i.e., non-multiplicative renormalization of $V_\mu^a$.
However, this mixing can be avoided,
since it is obviously equivalent to the
renormalization of the parameter $a_{\parallel l}$
with $\delta a_{\parallel l} = - 2 \epsilon a_{\parallel l}$.
}
in the covariant gauges even in this nonlinear system.

On the other hand,
the renormalization of the NG fields $\phi^i$ reads
$\delta\phi^i=
\delta Y^{(n)} / \delta K_i
= F^{(n)i}(\phi)$,
which implies that the parameterization of the
$G$-manifold is successively changed loop by loop
through point transformation.
It should be emphasized that
as explicit one-loop calculation shows\refmark{\HaradaYamawaki},
in the presence of the {\it propagating}
gauge fields
this function $F^{(n)i}(\phi)$ does not vanish
even if we use the dimensional regularization.
Thus there {\it is} a nontrivial renormalization on the
dimension-2 Lagrangian,
in sharp contrast to the
nonlinear model without
propagating gauge fields
(\cf\  Weinberg\refmark{\refWeinbergChPT}
and Appelquist-Bernard\refmark{\refAppBer}).


\chapter{Infrared Problem and Symmetry Breaking Mass Terms}

\section{Infrared Divergences}

Since we are treating massless Nambu-Goldstone (NG) fields $\phi $'s,
there generally appear infrared divergences which might invalidate the
formal discussions presented up to here.  But we now show that the
dimension-2 effective action $\Gamma ^{(n)}_2$,
as well as the $\Klinear$ terms in $\Gamma ^{(n)}_4$,
which is of our main concern in this paper, is in fact free from
the infrared divergences at least in the Landau gauge.

The NG fields $\phi ^i$ split into physical ones $\phi ^\a_\perp$
corresponding to the ``broken" generators
$X_\a \in {\cal G}-{\cal H}$, and unphysical ones $\phi ^a_{\para}$
corresponding to the ``unbroken" generators
$S_a \in {\cal H}$; namely,
$\phi ^iT_i=\phi ^\a_\perp X_\a + \phi ^a_{\para}S_a$.
They are in fact further decomposed into the $H$-irreducible
pieces $\phi ^{\a_k}_\perp \ (k=1,2,\cdots ,n)$
and $\phi ^{a_l}_{\para} \ (l=1,2,\cdots ,m)$ as explained in Sect.2.
The propagators of the physical NG fields $\phi ^{\a_k}_\perp$
are determined by the dimension-2 Lagrangian piece
$a_{\perp k}{\cal L}^{(k)}_\perp$ and given simply as the usual
massless ones:
$$
{\rm F.T.} i \bra{0} {\rm T}
\phi ^{\a_k}_\perp(x) \phi ^{\beta_k}_\perp(y) \ket{0} =
-{1\over p^2f_k^2} \delta _{\a_k \beta_k} \ ,
\eqn\eqPROPI
$$
with $f_k^2 \equiv  a_{\perp k}f_\pi^2$ and
F.T. denoting the Fourier transformation operation
$\int d^4(x-y) e^{ip(x-y)}$. The unphysical NG fields
$\phi ^{a_l}_{\para}$, on the other hand, generally mix with
the hidden gauge bosons $V_\mu ^{a_l}$ of the corresponding factor
group $H_l$.  Their propagators
are determined by the following quadratic pieces
of the Lagrangian $a_{\para l}{\cal L}_{\para}^{l}
+ {\cal L}_{\rm kin}(V_\mu)+{\cal L}_{\rm GF}$:
$$
{1\over 2}f_l^2(V_\mu ^{a_l}-\partial_\mu \phi ^{a_l}_{\para})^2
 -{1\over 4g_l^2} (\partial_\mu V_\nu ^{a_l}
-\partial_\nu V_\mu ^{a_l})^2
  +  B^{a_l} \partial^\mu V_\mu^{a_l}
  + {1\over2} \alpha_l g_l^2 B^{a_l} B^{a_l} \ ,
\ee
$$
with $f_l^2 \equiv  a_{\para l}f_\pi^2$.
Taking the inverse of the coefficient matrix of this quadratic form,
we find the following propagators:
$$
\eqalign{
{\rm F.T.} i \bra{0} {\rm T}
\phi ^{a_l}_\para(x) \phi ^{b_l}_\para(y) \ket{0} &=
   \delta _{a_lb_l}\Big[-{1\over p^2f_l^2}
+\a_l{g_l^2\over p^4}\Big]\ ,  \cr
{\rm F.T.} i \bra{0} {\rm T}
V_\mu ^{a_l}(x) \phi ^{b_l}_\para(y) \ket{0} &=
   \delta _{a_lb_l}\Big[-ip_\mu \a_l{g_l^2\over p^4}\Big]\ ,  \cr
{\rm F.T.} i \bra{0} {\rm T}
V_\mu ^{a_l}(x) V_\nu ^{b_l}(y) \ket{0} &=
   \delta _{a_lb_l}{g_l^2\over p^2-g_l^2f_l^2}
      \Big[g_{\mu \nu }-{p_\mu p_\nu \over p^2}
        \Big(1-\a_l+\a_l{g_l^2 f_l^2\over p^2}\Big)\Big]\ .  \cr
 }\eqn\eqPROPII
$$
These are the propagators in a general covariant gauge with gauge
parameter $\a_l$.
In the case of non-Landau gauge $\a_l\not=0$
there exists
a {\it massless dipole} $1/p^4$ in these
$\phi _\para$-$\phi _\para$ and $V_\mu $-$\phi _\para$ propagators,
which are
not well-defined as they stand
even in the sense of distribution just like the massless
boson propagators in two dimension.
[The massless dipole
$p_\mu p_\nu /p^4$ in the $V_\mu $-$V_\nu $ propagator,
on the other hand, can be well-defined by the presence of
$p_\mu p_\nu $ in the numerator.]

So let us consider the Landau gauge case $\a_l=0$ first.
Then there appear no massless dipoles and no
$V_\mu $-$\phi _\para$ transition propagators.
We note the followings.
As far as the infrared behavior is concerned, we can rewrite the
vector propagator into
$$
\eqalign{
&\Big(g_{\mu \nu }-{p_\mu p_\nu \over p^2}\Big)
{g_l^2\over p^2-g_l^2f_l^2}  \cr
 & \ \qquad \ \ =
\Big(g_{\mu \nu }-{p_\mu p_\nu \over p^2}\Big) (-{1\over f_l^2})
\Big[ 1 + \Big({p^2\over g_l^2f_l^2}\Big)
+ \Big({p^2\over g_l^2f_l^2}\Big)^2
+ \cdots  \Big]
\ . \cr}
\ee
$$
This \hfill expansion \hfill just \hfill corresponds
\hfill to \hfill the \hfill perturbation \hfill expansion
\hfill in \hfill which \hfill we \hfill treat
\hfill the \hfill dimension-4 \hfill Lagrangian
\hfill
${\cal L}_{\rm kin}(V_\mu)$
\hfill as \hfill perturbation \hfill interactions
\hfill and \hfill use \hfill the \hfill propagator
\hfill
$-(g_{\mu \nu }-{p_\mu p_\nu /p^2})/f_l^2$
\hfill determined \hfill solely \hfill by the dimension-2
Lagrangian $a_{\para l}{\cal L}_{\para}^{l}$.
This type of perturbation
expansion is clearly not valid from the ultraviolet view point, but
provides us with completely legitimate one for our present purpose
considering the infrared behavior of our Green functions. We mean that
we use this expansion only in the vicinity of $p_\mu =0$ of the loop
momentum $p_\mu $ in the relevant Feynman integrands.

With this
expansion, all the propagators of $\phi _{\perp}$, $\phi _{\para}$ and
$V_\mu $ particles, \eqPROPI\ and \eqPROPII, are now
determined by the dimension-2 Lagrangians and hence
are proportional to the inverse power of the decay constant,
$1/f_\pi^2$. Moreover, we can rescale the FP anti-ghost field
$\bar C$ into $f_\pi^2\bar C$ so that the FP Lagrangian become
${\cal L}_{\rm FP}= i f_\pi^2 \bar{C}^a \partial^\mu D_\mu C^a $,
then the FP ghost propagator also becomes proportional to
the inverse power $1/f_\pi^2$.
We can now count the power of $f_\pi^2$ for a general
Feynman diagram contributing to $\Gamma $.
If we use only the vertices of the dimension-2
Lagrangians or the FP ghost one ${\cal L}_{\rm FP}$,
which are all proportional to $f_\pi^2$, the counting of the
inverse power of $f_\pi^2$ is the same as that of Planck constant
$\hbar$, and so we get $(1/f_\pi^2)^{(L-1)}$ for any $L$-loop diagrams.
But if we use the vertices coming from the dimension-4 Lagrangian
${\cal L}_{\rm kin}(V_\mu)$ or the BRS source terms
$K_i \dBRS \phi^i + {\mib K}_A^\mu
\dBRS \VA_\mu^A + {\mib L}_A \dBRS {\mib C}^A$,
which have no power of $f_\pi^2$, we lose
a power of  $f_\pi^2$ for each of such vertices. Therefore, a general
$L$-loop Feynman diagram possessing $V_4$ vertices of
${\cal L}_{\rm kin}(V_\mu)$ and $K$ BRS source vertices,
yields an amplitude proportional to%
\foot{
This counting is similar to that in the nonlinear
chiral Lagrangian\refmark{\refWeinbergChPT}.
}
$$
\Big({1\over f_\pi^2}\Big)^{(L-1+V_4+K)} \ .
\ee
$$
But this implies the following: First the dimension-2 effective
action $\Gamma ^{(n)}_2$, which is proportional to $f_\pi^2$,
receives no
loop corrections at all, since the power $L-1+V_4+K$ is non-negative
because $L\geq 1,\ V_4,K\geq 0$.
Moreover, the $\Klinear$ terms in
the dimension-4 effective action $\Gamma ^{(n)}_4$,
which is of zero-th power
term in $f_\pi^2$, also have no contributions since the BRS source
vertex should be contained once there, $K=1$, so $L-1+V_4+K\geq 1$.
[Note that all these are concerned with only the infrared
contributions. There are actually non-zero loop contributions to
$\Gamma ^{(n)}_2$ as well as to the $\Klinear$ terms in
$\Gamma ^{(n)}_4$
coming from the ultraviolet region, for which the above counting of
the powers of $f_\pi^2$ breaks down.] This finishes the proof of the
absence of the infrared divergences in
$\Gamma ^{(n)}_2$ and the $\Klinear$ terms in $\Gamma ^{(n)}_4$ in
the case of
Landau gauge.

In the non-Landau gauge case, we should properly define the massless
dipole propagator $1/p^4$.  For instance, it can be defined by
introducing a small infrared cutoff for the time-component $p_0$ as
Nakanishi did long time ago in QED\refmark{\refNakanishi}.
Then, as far as the infrared cutoff is
kept finite, there appear no infrared divergences of the dipole
propagator origin. So the above argument for the Landau gauge case
applies also to this non-Landau gauge case and shows that there appear
no infrared divergences in $\Gamma ^{(n)}_2$ and
the $\Klinear$ terms in $\Gamma ^{(n)}_4$. But there are non-trivial
problems whether the theory recovers the Lorentz invariance or not
in the limit when the infrared cutoff goes to zero,
or whether it remains well-defined in that limit.
We here do not pursue these problems any further.

\section{Symmetry Breaking Mass Terms}

Finally, we make a comment on what happens when there exist symmetry
breaking mass terms of NG fields $\phi $.
Such mass terms may appear when there are explicit $G$-symmetry
breaking as in the chiral symmetry in QCD case, or when we want to
regularize the infrared divergences as a technical device.

When there exist such a mass term of $\phi $, we introduce the
following
BRS source term corresponding to its BRS transformation:
$$
{\cal L}_{\rm mass}
= f_\pi^2 \cdot m^2 f(\phi ) + M \dBRS f(\phi )\ ,
\eqn\eqMASS
$$
where the function $f(\phi )$ is $\phi^2/2 \equiv f^{(0)}(\phi)$
at the tree level and is
generally dimension 0 function containing no derivatives.
For simplicity we assume that the mass terms preserves
the $H_{\rm diag}$ symmetry. We assign dimensions
2 and 4 to $m^2$ and the BRS source $M$, respectively.
In the presence of this mass term, the WT identity for the effective
action becomes
$$
  \Gamma \ast \Gamma = \mtil^2{\delta \Gamma \over \delta M}\ ,
\ee
$$
with $\mtil^2\equiv m^2 f_\pi^2$ (which is still dimension 2 in
our counting).
Then our renormalization equations \DimzeroWTB\ -- \DimfourWTB\ are
changed into
$$
\eqalignno{
  & \dBRS \Gamma_0^{(n)} = 0 \ ,
&\eqname{\zeroWTB} \cr
  & \dBRS \Gamma_2^{(n)} + \delta'_\Gamma S_2
 =  \mtil^2{\delta \Gamma ^{(n)}_4\over \delta M}\ ,
&\eqname{\twoWTB} \cr
  & \dBRS \Gamma_4^{(n)} + \delta'_\Gamma S_4
  + {\delta \Gamma_6^{(n)} \over \delta \Kex}
    {\delta S_2 \over \delta \Phi}
 =  \mtil^2{\delta \Gamma ^{(n)}_6\over \delta M}\ ,
&\eqname{\fourWTB}
}
$$
respectively. The first equation is the same as before, so that it
still leads to $\Gamma ^{(n)}_0=0$.  The third equation \fourWTB,
whose
$\Klinear$ terms determined the form of $\delta'_\Gamma$ previously,
now has a non-vanishing right hand side. But, fortunately, it does
not contributes to the relevant $\Klinear$ terms since
$\delta \Gamma ^{(n)}_6/\delta M$ is of dimension $6-4=2$ and cannot
contain $\Kex$
of dimension 3 or 4. Therefore $\delta'_\Gamma$ is determined in the
same form as before as given in the Appendix A.
[There we need the assumption
that the mass terms respect the $H_{\rm diag}$ symmetry.]
Finally consider the second equation \twoWTB\ with $\delta'_\Gamma$
thus determined. Repeating the same argument as performed before in
the subsection A.2 in the Appendix A, and noting that the dimension-2
tree action $S_2$ now contains BRS- (and $H_{\rm local}$-)
non-invariant mass term
$\mtil^2 f^{(0)}(\phi)= \mtil^2\phi^2/2$, we easily
find
$$
  \deltaGam S_2 =   \dBRS
  \left({
    \hat{F}^{(n)} S_2 + \alpha_l^{(n)} V_\mu^a
    {\delta \over \delta V_\mu^a} S_2
  }\right)
  -\mtil^2 \hat{F}^{(n)}\big(\dBRS f^{(0)} \big)
  +\mtil^2 \beta_l^{(n)} C^a \hat{W}_a f^{(0)} \ ,
\ee
$$
so that \twoWTB\ turns out to give
$$
  \dBRS \Big[ \Gamma_2^{(n)} +  \Big({
    \hat{F}^{(n)} S_2 + \alpha_l^{(n)} V_\mu^a
    {\delta \over \delta V_\mu^a} S_2
  }\Big) \Big]
 =
  \mtil^2 \big(\hat{F}^{(n)}  - \beta_l^{(n)}
     C^a {\delta \over \delta C^a}\big) \dBRS f^{(0)}
  + \mtil^2{\delta \Gamma ^{(n)}_4\over \delta M}\ ,
\eqn\AAAA
$$
where the integral $\int d^4x f^{(0)}(\phi)$ is denoted simply by
$f^{(0)}$.
The BRS operand $\Gamma _2^{(n)} + \cdots $ in the left hand side is
dimension-2 quantity and has generally the form
$$
A_2[\phi,\VA] + \mtil^2 A_0[\phi]\ ,
\eqn\eqDDD
$$
where $A_2$ is a dimension-2 functional with dimensions given by
derivatives and/or gauge fields, and $A_0$ is a dimension-0 functional
of $\phi $ with no derivatives. [To avoid any confusion, we should
note that
\eqDDD\ is {\it not} a Taylor expansion in $\mtil^2$.  We have written
$\mtil^2$ explicitly in front of the second term since it carries the
dimension of that term, but have not written explicitly any $m^2$'s
in dimensionless form like $\ln (m^2/\mu ^2)$, which may still
appear both in $A_2$ and $A_0$.]  Let us call the dimension carried
by the derivatives and fields alone {\it genuine-dimension}. Then the
first term $A_2$ is a genuine-dimension 2
term and the second $\mtil^2A_0$
a genuine-dimension 0 term. Two terms with different
genuine-dimensions are of course mutually independent. Note that
all the terms in the right hand side
of \AAAA\ are of genuine-dimension 0, so we see that \AAAA\
gives the following two equations for the genuine-dimension 2
and 0 parts, respectively:
$$
\eqalign{
\dBRS A_2 &= 0\ , \cr
\dBRS A_0 &=
   \big(\hat{F}^{(n)}  - \beta_l^{(n)}
     C^a {\delta \over \delta C^a}\big) \dBRS f^{(0)}
   + {\delta \Gamma ^{(n)}_4\over \delta M}\ . \cr
}\ee
$$
The first equation says that $A_2$ is given by
a gauge-invariant functional $A_{2\GI} [\phi,\VA] $ just as in the
case of no mass term. The second equation does not constrain the
form of $A_0$
but determines the $M$-dependence of $\Gamma ^{(n)}_4$; namely,
writing the dimension-0 functional $A_0[\phi ]$ as
$-\int d^4x f^{(n)}(\phi )$ generically, we find that
our effective action $\Gamma ^{(n)}$
contains the following additional terms in the presence of mass terms.
$$
\int d^4x \Big[- \mtil^2 f^{(n)}(\phi ) - M \dBRS f^{(n)}(\phi ) \Big]
 - M \big( \hat{F}^{(n)}  - \beta_l^{(n)}
     C^a {\delta \over \delta C^a}\big) \dBRS f^{(0)}
\ .
\ee
$$
But, when writing the solution $\Gamma^{(n)}$ in the form
$- S \ast Y^{(n)}$ as in \Solution, we should note that
$- S \ast Y^{(n)}$  now contains an additional piece
$ -(M\dBRS f^{(0)}) \ast Y^{(n)}$ which exactly yields the second
term $ - M \big( \hat{F}^{(n)}  - \beta_l^{(n)}
     C^a (\delta / \delta C^a)\big) \dBRS f^{(0)} $.
Thus the solution of the renormalization equations \zeroWTB\ --
\fourWTB\ turns out to be given by
$$
  \Gamma_2^{(n)} +
  \left. \Gamma_4^{(n)} \right\vert_{\Klinear}
  =
  A_{2\GI}^{(n)} [\phi,\VA] - S \ast Y^{(n)} +
\int d^4x \Big[- \mtil^2 f^{(n)}(\phi ) - M \dBRS f^{(n)}(\phi ) \Big]
\ ,
\ee
$$
with the understanding that the ``$\Klinear$ term" here contains
$M$-linear terms also. The additional term
which newly appears compared with the previous
solution \Solution\ is only the last two terms.
These two terms can be canceled by renormalizing the mass function
$f(\phi )$ in the mass term \eqMASS\ as
$$
\big(f(\phi )\big)_n = {1\over 2}\phi ^2 + f^{(1)}(\phi ) + \cdots
 + f^{(n)}(\phi ) \ .
\ee
$$
This implies that we can carry out our renormalization procedure
even in the presence of $G$-symmetry breaking mass terms and,
in particular, our main Proposition in Sect.3 concerning the
(genuine-) dimension-2 Lagrangian remains intact.


\chapter{Summary and Discussion}

We have shown in the covariant gauges
that our tree-level dimension-2 action
$S_2$, \SfourAction,
if written in terms of renormalized parameters and fields,
already gives the exact action $\Gamma_2$
including all the loop effects.
The proof was done for $\Gglobal\times\Hlocal$ model,
with $G$ and $H$ ($\subset G$)
being arbitrary compact groups.

Our conclusion in this paper remains unaltered
even if the action $S$ contains other dimension-4 or higher terms,
as far as they respect the symmetry.
This is because we needed just
$\left({S \ast \Gamma}\right)_2$
and $\left({S \ast \Gamma}\right)_4 \vert_{\Klinear}$
parts in the WT identity
to which only
$S_2$ and $\Klinear$ part
of $S_4$ can contribute.

Our model includes the chiral model
with
$G=U(N)_{\rm L}\times U(N)_{\rm R}$
and $H=U(N)_{\rm V}$.
The case of chiral model was worked out more explicitly
in a separate article\refmark{\HaradaKugoYamawakiPRL}.
When we regard this chiral model
as a low energy effective theory of QCD,
we must
take account of
the anomaly
and the corresponding
Wess-Zumino-Witten term $\Gamma_{\rm WZW}$.
The WT identity
then reads
$\Gamma\ast\Gamma=({\rm anomaly})$.
However,
the RHS
is saturated already at the tree level
in this effective Lagrangian,
and hence the WT identity
at loop levels,
which we need,
remains the same as before.
The WZW term $\Gamma_{\rm WZW}$ or
any other intrinsic-parity-odd terms\refmark{\FKTUY} in $S$
are of dimension-4 or higher and hence
do not change our conclusion as explained above.

In the chiral model $S_2$ takes a simple form
$\int d^4 x \left( {\cal L}_{\rm A} + a {\cal L}_{\rm V} \right)$,
which
(in particular the ${\cal L}_{\rm V}$ part) implies that
the previously derived
relation\refmark{\BandoKugoYamawakiNP,\BandoKugoYamawakiPTP}
$$
{g_V (p^2) \over
g_{V\pi\pi}(p^2, p_{\pi_1}^2\!\!=\! p_{\pi_2}^2\!\!=\!0)}
\bigg\vert_{p^2=0} = 2f_\pi^2
\eqn\eqLET
$$
{\it is} actually an exact low energy theorem valid at any loop
order. Of course, this theorem concerns off-shell quantities at
$p^2=0$ of the vector field momentum $p$,
and hence is not physical as it stands.
However, suppose that the vector mass
$m_V^2=ag^2f_\pi^2$ is sufficiently small compared
with the characteristic energy scale $\Lambda^2$ of the system,
which is customarily taken as $\Lambda^2\sim 16\pi^2 f_\pi^2$.
Then we expect that the on-shell value of
$g_V/g_{V\pi\pi}$ at $p^2=m_V^2$ can deviate
from the LHS of \eqLET\ only by a quantity of order
$m_V^2/\Lambda^2\sim ag^2/16\pi^2$, since the contributions of the
dimension-4 or higher terms in the effective
action $\Gamma$ (again representing all the loop effects) are
suppressed by a factor of $p^2/\Lambda^2$ at least.
Therefore as far as the vector mass is light,
namely, when either $a$ or $g^2/16\pi^2$ is small,
our theorem is truly a physical one
\foot{
It is interesting to note
that the ``vector limit''\refmark{\Georgi}
realizes this light vector
meson limit.}.

In the actual world of QCD, the $\rho$-meson mass
is not so light ($ag^2/16\pi^2\sim 1/2$) so that the situation
becomes a bit obscure. Nevertheless,
the fact that the KSRF (I) relation
$g_{\rho}/g_{\rho\pi\pi}=2f_\pi^2$
holds on the $\rho$ mass shell with good accuracy
strongly suggests that the $\rho$-meson is
the hidden gauge field
and
{\it the KSRF (I) relation is a physical manifestation of our
low energy theorem}.

In this connection we should comment on the gauge choice.
In the covariant gauges which we adopted here,
the $\Gglobal$ and $\Hlocal$ BRS symmetries
are separately preserved.
Accordingly,
the $\Hgf$ field is multiplicatively
renormalized,
and the above (off-shell) low energy theorem \eqLET\ holds.
However, if we adopt $R_\xi$-gauges
(other than Landau gauge),
these properties are violated;
for instance, $\phi\partial_\mu \phi$ or the external gauge field
${\cal V}_\mu$ gets mixed in the renormalization of $V_\mu$,
and our off-shell low energy theorem \eqLET\ is violated.
This implies that the $\Hgf$ field in the $R_\xi$ gauge
generally does not give a smooth off-shell extrapolation;
indeed, in $R_\xi$ gauge with gauge parameter
$\alpha \equiv 1/\xi$, the correction
to $g_{\rho}/g_{\rho\pi\pi}$ by
the extrapolation from $p^2=m_\rho^2$ to $p^2=0$ is seen
to have a part proportional to $\alpha g^2/16\pi^2$, which diverges
when $\alpha$ becomes very large.
Thus, in particular, the unitary gauge%
\foot{In the unitary gauge our hidden local symmetry
Lagrangian coincides with the
Weinberg's old Lagrangian
for the $\rho$ meson\refmark{\WeinbergRho}.
},
which corresponds to
$\alpha \rightarrow \infty$,
gives an ill-defined off-shell field.

\ACK
We would like to thank Volodya Miransky and
Masaharu Tanabashi for stimulating discussions.
T.~K. is supported in part by the Grant-in-Aid for
 Scientific Research (\#04640292) from the Ministry of Education,
Science and Culture.
K.~Y. is supported in part by the Takeda Science Foundation
and the Ishida Foundation,
and also by the International Collaboration Program
of the Japan Society for Promotion of Science.

\endpage


\APPENDIX{A}{A.\  Solution to the Renormalization
Equations \DimtwoWTB \ and \DimfourWTB}

In this Appendix, we prove that the general solution
to \DimtwoWTB\ and \DimfourWTB\ is given
in the form \Solution.
We first solve \DimfourWTB\ which determines the form
of the $\Kex$-linear term in $\Gamma_4^{(n)}$,
or equivalently, the form of $\delta'_\Gamma$.
Then next we solve \DimtwoWTB\ using the obtained
$\delta'_\Gamma$ and determine the form of $\Gamma_2^{(n)}$.

\section{ Solving Eq.\DimfourWTB}

To determine the form of $\deltaGam$,
we in fact use the information only of the $\Kex$-linear terms
of the \DimfourWTB.
To the $\Kex$-linear terms of \DimfourWTB,
only the $\Kex$-linear terms in $\Gamma_4^{(n)}$
and the $\Kex$-quadratic terms in $\Gamma_6^{(n)}$ can contribute.
Taking account of the ghost numbers and dimension,
we write the general forms of them as
$$
  \left. \Gamma_4^{(n)} \right\vert_{\Klinear}
  =
  K_i \deltaGam\phi^i + K_a^\mu \deltaGam V_\mu^a
  + L_a \deltaGam C^a
\eqn{\Gamfour}
$$
with
$$
\eqalignno{
  \deltaGam \phi^i
&=
  {\mib C}^A R_A^i (\phi) ,
&\eqname{\Phitrans}
\cr
  \deltaGam V_\mu^a
&=
  {G_b}^a (\phi) \partial_\mu C^b +
  \left[{
    {G_{bi}}^a (\phi) \partial_\mu \phi^i
    + {H_{bc}}^a (\phi) V_\mu^c
    + {{\cal H}_{bi}}^a (\phi) {\cal V}_\mu^i
  }\right]
  C^b ,
&\eqname{\Vtrans}
\cr
  \deltaGam C^a
&=
  - {1\over2} C^b C^c {R_{[bc]}}^a (\phi) ,
&\eqname{\Ctrans}
\cr
  \left. \Gamma_6^{(n)} \right\vert_{\Kquadratic}
&=
  {1\over f_\pi^2} F_{[ab][cd]} (\phi) K^{a\mu} K_\mu^b C^c C^d ,
&\eqname{\Gamsix}
}
$$
where $R_A^i$, $G_b^a$, ${G_{bi}}^a$, ${H_{bc}}^a$,
${{\cal H}_{bi}}^a$, ${R_{[bc]}}^a$ and $F_{[ab][cd]}$
are dimension-0 functions of $\phi^i$ (without derivatives)
carrying the specified group index structures.
(The notation $[ab]$ means anti-symmetry under $a \leftrightarrow b$.)
They are arbitrary functions at this stage.
Note that the BRS source terms
${\cal K}_i^\mu\deltaGam{\cal V}_\mu^i$
and ${\cal L}_i\deltaGam{\cal C}^i$ did not appear in \Gamfour.
This is because the external $\Gglobal$-gauge fields ${\cal V}_\mu^i$
as well as their ghosts ${\cal C}^i$ are not quantized
but merely c-number fields in our system,
and therefore their BRS sources ${\cal K}_i^\mu$ and ${\cal L}_i$
appear only in the tree action.
So we have $\deltaGam{\cal V}_\mu^i=\deltaGam{\cal C}^i=0$.
Note also that only the $\Hlocal $ ghosts $C^a$ appear in
\Vtrans\ -- \Gamsix.
This is because the ghost numbers of $\Hlocal $ ghosts $C^a$
and of $\Gglobal $ ghosts ${\cal C}^i$ are in fact {\it separately}
conserved in the present system of covariant gauge
\GFFPlagrangian, and the BRS sources $K_a^\mu$ and $L_a$ carries
the $\Hlocal $ ghost number $-1$ and $-2$, respectively.
[Note also that $L_a$'s with the index $a$ corresponding to
$U(1)$ factor groups $H_l$ are absent since $\dBRS C^a=0$.]
On the contrary, the BRS source $K_i$ for $\phi^i$ cannot be
assigned any definite separate ghost number,
so that both ghosts $C^a$ and ${\cal C}^i$ appear in \Phitrans:
$$
  {\mib C}^A R_A^i(\phi) \equiv C^a R_a^i(\phi)
  + {\cal C}^j {\cal R}_j^i(\phi).
\eqno\eq
$$

Picking up the \Kex-linear terms of eq.\DimfourWTB\
by inserting \Gamfour\ and \Gamsix, we find
$$
\eqalign{
&{}
  -K_i \{\dBRS,\deltaGam\} \phi^i
  + L_a \{\dBRS,\deltaGam\} C^a
\cr
&{} \qquad\quad
  - K_a^\mu
  \bigg({
    \{\dBRS,\deltaGam\} V_\mu^a
    + 2 F_{[ab][cd]} a_{\parallel l}
      (V_\mu^b - \Apara_\mu^b) C^c C^d
  }\bigg)
  =0 .
}
\eqn{\Klinterm}
$$
In the last term we have used
$\delta S_2 / \delta V_\mu^b =
 a_{\parallel l} ( V_\mu^b - \Apara_\mu^b )$
obtained from \Lparallel, and accordingly the index $l$
of $a_{\parallel l}$ should be understood to refer to
the factor group $H_l$ to which the $\Hlocal $-group
index $b$ of $V_\mu^b$ belongs.
Since \Klinterm\ holds as an identity,
the terms proportional to $K_i$, $L_a$ and $K_a^\mu$ have to
vanish separately.
We examine those in the order
i) $L_a$ term,
ii) $K_i$ term and
iii) $K_a^\mu$ term.

The $L_a$ term demands $\{\dBRS,\deltaGam\}C^a=0$,
the two terms of which are calculated using \PhitransB,
\CtransB \ and \Ctrans \ to be
$$
\eqalignno{
  \dBRS ( \deltaGam C^a )
&=
  -{1\over2} \dBRS
  \left({
    C^b C^c {R_{[bc]}}^a (\phi)
  }\right)
& \
\cr
&=
  -{1\over2}
  \left[{
    {1\over2} C^b \left({ C \times C}\right)^c {R_{[bc]}}^a
    - {1\over2} \left({ C \times C}\right)^b C^c {R_{[bc]}}^a
    + C^b C^c {\mib C}^A (\hat{\mib W}_A {R_{[bc]}}^a)
  }\right] ,
&\eqname{\LatransA}
\cr
  \deltaGam ( \dBRS C^a )
&=
  -{1\over2} \deltaGam
  \left({
    C^b C^c {f_{bc}}^a
  }\right)
& \
\cr
&=
  {1\over4}
  \left[{
    C^d C^e {R_{[de]}}^b C^c {f_{bc}}^a
    - C^b C^d C^e {R_{[de]}}^c {f_{bc}}^a
  }\right] ,
&\eqname{\LatransB}
}
$$
with notation
$\left({ C \times C}\right)^a \equiv C^b C^c {f_{bc}}^a$.
Again \LatransA$+$\LatransB$=0$ is an identity in the field variables.
Since the $\Gglobal $-ghost fields ${\cal C}^i$ appear
only in the last term in \LatransA \ in the form
$C^b C^c {\cal C}^i ( \hat{{\cal W}}_i {R_{[bc]}}^a )$
[recall that
$ {\mib C}^A \hat{\mib W}_A = C^a \hat{W}_a +
{\cal C}^i \hat{{\cal W}}_i $ ],
that term should vanish by itself and so
$$
  \hat{{\cal W}}_i {R_{[bc]}}^a (\phi) = 0 .
\eqno\eq
$$
As explained in \Solzero, this holds only when ${R_{[bc]}}^a(\phi)$
is a $\phi$-independent constant.
But such a constant, which carries the index structure ${R_{[bc]}}^a$
under the (unbroken) $\Hdiag $ group and satisfies antisymmetry
under $b \leftrightarrow c$, is only the structure constant
${f_{bc}}^a$,
[Recall that $L_a$ is absent for $U(1)$ factor groups and
so the index $a$ here belongs to a certain simple factor group.]
and so we have
$$
  {R_{[bc]}}^a (\phi) = \beta_l {f_{bc}}^a .
\eqn{\SolR}
$$
Here the proportionality constant $\beta_l$ may depends
on the factor group $H_l$ to which the indices $a$, $b$, $c$ belong.
[For later convenience, we take $\beta_l=0$ for $U(1)$ factor groups
$H_l$ as convention although \SolR\ is zero in any case.]
Namely, at this stage we find
$$
  \deltaGam C^a = - {1\over2} \beta_l
  \left({ C \times C }\right)^a
  = \beta_l \dBRS C^a ,
\eqn{\SolCtrans}
$$
so that $ \{ \dBRS , \deltaGam \} C^a = 0 $ is now clear
from the nilpotency of $\dBRS$, $\left({\dBRS}\right)^2=0$.

Next consider the $K_i$ term demanding
$ \{ \dBRS ,\deltaGam \} \phi^i = 0 $.
Using \PhitransB, \CtransB, \Phitrans\ and \SolCtrans, we find
$$
\eqalign{
  \left\{{ \dBRS , \deltaGam }\right\} \phi^i
&=
  {\mib C}^A {\mib C}^B
  \left[{
    \hat{R}_A {\mib W}_B^i + \hat{\mib W}_A R_B^i
    - {1\over2} {f_{AB}}^C R_C^i
  }\right]
\cr
& \qquad
  - {\beta_l \over 2} C^a C^b {f_{ab}}^c W_c^i ,
}
\eqn{\Kitrans}
$$
with notation
$ \hat{R}_A = R_A^i (\phi) {\partial \over \partial \phi^i} $.
Taking account of the anti-commutativity of the ghosts,
vanishingness of \Kitrans\ means the relation
$$
  \big[{ \hat{\mib W}_A , \hat{R}_B }\big]
  - \big[{ \hat{\mib W}_B , \hat{R}_A }\big]
  =
  {f_{AB}}^C \hat{R}_C
  + \delta_A^a \delta_B^b {f_{ab}}^c \hat{W}_c .
\eqno\eq
$$
This should be solved with respect to $\hat{R}_A$.
The last term, which contributes only when the indices $A$ and $B$
are $\Hlocal $ ones $a$ and $b$,
can easily be eliminated by shifting the $\hat{R}_A$ operator as
$$
  \hat{R}_A = \hat{R}'_A + \delta_A^a \beta_l \hat{W}_a .
\eqn{\eqAXV}
$$
Indeed, then using
$ \big[{ \hat{W}_a , \hat{W}_b }\big] = {f_{ab}}^c \hat{W}_c $
in \RelOfAlgebra, we find the following homogeneous equation
for $\hat{R}'_A$;
$$
  \big[{ \hat{\mib W}_A , \hat{R}'_B }\big]
  - \big[{ \hat{\mib W}_B , \hat{R}'_A }\big]
  =
  {f_{AB}}^C \hat{R}'_C .
\eqn{\RelOfRd}
$$
Clearly from the algebra
$ \big[{ \hat{\mib W}_A , \hat{\mib W}_B }\big] =
{f_{AB}}^C \hat{\mib W}_C $
in \RelOfAlgebra, $\hat{R}'_A$ of the form
$$
  \hat{R}'_A = \big[{ \hat{\mib W}_A , \hat{F} }\big] ,
  \qquad
  \hat{F} \equiv F^i (\phi) {\partial \over \partial \phi^i}
\eqn{\SolOfRd}
$$
with some function $F^i(\phi)$, satisfies \RelOfRd.
But it is less trivial whether any solution to \RelOfRd \ can be
written in the form \SolOfRd.
This is proved in Appendix B.

Finally consider the $K_a^\mu$ term.
First of all we should note that $K_a^\mu$ with index $a$
corresponding to any $U(1)$ factor group in $\Hlocal $
does not appear in $\Gamma^{(n)}$.
This is because $K_a^\mu$ is contained in the original action $S$
only in the form $ K_a^\mu \partial_\mu C^a $
for the $U(1)$ group index $a$.
Since the $U(1)$ ghost $C^a$ is free in this covariant gauge,
$K_a^\mu \partial_\mu C^a$ itself is like a c-number source
and never appears in $\Gamma^{(n)}$.
So we can assume the group index $a$ of $K_a^\mu$ belongs
to a certain simple factor group $H_l$ henceforth.
Note also that the index $b$ of $C^b$ in \Vtrans\
also belongs to the same simple factor group $H_l$,
since the ghost number is in fact conserved separately
for each factor group in this covariant gauge.

Let us now analyze the vanishingness condition of the $K_a^\mu$ term
in \Klinterm:
$$
  \left\{ { \dBRS , \deltaGam }\right\} V_\mu^a
  + 2 F_{[ab][cd]} a_{\parallel l}
  \left({ V_\mu^b - \Apara_\mu^b }\right)
  C^c C^d =0.
\eqn{\KlinV}
$$
Using \Vtrans, we calculate
$$
\eqalign{
  \dBRS \left({ \deltaGam V_\mu^a }\right)
&=
  {\mib C}^A \left({ \hat{\mib W}_A G_b^a }\right) \partial_\mu C^b
\cr
& \qquad
  +
  \Bigl[
    {\mib C}^A \hat{\mib W}_A
      \left({ {G_{bi}}^a \partial_\mu \phi^i }\right)
    + {G_{bi}}^a {\mib W}_A^i \partial_\mu {\mib C}^A
\cr
& \qquad\quad
    + {\mib C}^A \left({ \hat{\mib W}_A {H_{bc}}^a }\right) V_\mu^c
    + {H_{bc}}^a D_\mu C^b
\cr
& \qquad\quad
    + {\mib C}^A \left({ \hat{\mib W}_A {{\cal H}_{bi}}^a }\right)
      {\cal V}_\mu^i
    + {{\cal H}_{bi}}^a D_\mu {\cal C}^i
  \Bigr]
  C^b
\cr
& \qquad
  +
  \left[{
    {G_{bi}}^a \partial_\mu \phi^i
    + {H_{bc}}^a V_\mu^c
    + {{\cal H}_{bi}}^a {\cal V}_\mu^i
  }\right]
  \Big({
    -{1\over2} C \times C
  }\Big)^b .
}
\eqn{\KlinVA}
$$
Noting that the $\Gglobal $ ghosts ${\cal C}^i$ do not appear in
$\deltaGam ( \dBRS V_\mu^a )$,
we see that the terms containing any ${\cal C}$ in \KlinVA\
should cancel among them in order for \KlinV\ to hold.
There are various types of such terms,
each of which gives the constraint on the coefficient functions
in \Vtrans:
$$
\eqalignno{
& {\cal C}^i \partial_\mu C^b \quad :
  \qquad\quad
  \hat{{\cal W}}_i G_b^a (\phi) = 0 ,
&\eqname{\CdelCterm}
\cr
& {\cal C}^j C^b \partial_\mu \phi^i \quad :
  \qquad
  \hat{{\cal W}}_j \left({ {G_{bi}}^a(\phi)
  \partial_\mu \phi^i }\right) = 0 ,
&\eqname{\CCdelPhiterm}
\cr
& \partial_\mu {\cal C}^i C_b \quad :
  \qquad\quad
  {G_{bj}}^a(\phi) {\cal W}_i^j(\phi) + {{\cal H}_{bi}}^a(\phi) = 0 ,
&\eqname{\delCCterm}
\cr
& {\cal C}^i C^b V_\mu^c \quad :
  \qquad\quad
  \hat{{\cal W}}_i {H_{bc}}^a (\phi) = 0 ,
&\eqname{\CCVterm}
\cr
& {\cal C}^i C^b {\cal V}_\mu^j \quad :
  \qquad\quad
  \hat{\cal W}_i {{\cal H}_{bj}}^a (\phi)
  + {{\cal H}_{bk}}^a(\phi) {f_{ji}}^k = 0 .
&\eqname{\CCcalVterm}
}
$$

Eq.\CdelCterm \ says that $G_b^a(\phi)$ is
$\phi$-independent constant by \Solzero.
By $\Hdiag $-covariance and the separate ghost number conservation
for each factor group $H_l$,
such a constant must be proportional to $\delta_b^a$:
$$
  G_b^a (\phi) = \alpha'_l \delta_b^a ,
\eqno\eq
$$
where the proportionality constant $\alpha'_l$ may depend
on the simple factor group $H_l$ to which the index $a$ belongs.

Eq.\CCdelPhiterm\  does not say that ${G_{bi}}^a \partial_\mu \phi^i$
is constant since it contains derivative $\partial_\mu\phi^i$,
but says it is $\Gglobal $-invariant.
We know that the only $\Gglobal $-invariant containing
the first order derivative $\partial_\mu\phi^i$
is given by
$\alpha_\mu^i (\phi) \equiv {2\over i}
 \tr \big({ T^i \partial_\mu \xi \cdot \xi^{\dag} }\big)$,
or, separating the $H$ and $G/H$ generator parts, by
$\alpha_\mu^a (\phi) \equiv {2\over i}
 \tr \big({ S^a \partial_\mu \xi \cdot \xi^{\dag} }\big)$ and
$\alpha_{\mu\perp}^\alpha (\phi) \equiv {2\over i}
 \tr \big({ X^\alpha \partial_\mu \xi \cdot \xi^{\dag} }\big)$.
Therefore the $\Gglobal$-invariant
${G_{bi}}^a (\phi) \partial_\mu \phi^i$ must be a linear combination
of them:
$$
  {G_{bi}}^a (\phi) \partial_\mu \phi^i
  =
  {g_{bc}}^a \alpha_\mu^c (\phi)
   +  {g^{\perp}_{b\alpha}}^a \alpm^\alpha (\phi) \ .
\eqn{\Gbiaterm}
$$
Since the indices $a$ and $b$ belong
to the same simple factor group $H_l$ as noted in the above,
the $\Hdiag $-covariance demands the coefficient ${g_{bc}}^a$
corresponding to the $H$ group index $c$ to take the form
$$
  {g_{bc}}^a =
  \gamma_l {f_{bc}}^a
  + \tilde{\gamma}_l {d_{bc}}^a
  + \tilde{\tilde{\gamma}}_{lu_1} \delta_b^a \delta_c^{u_1} \ ,
\eqno\eq
$$
where
$d_{abc} \equiv 2 \tr \left({ S_a \{ S_b , S_c \} }\right)$
for $S_a$, $S_b$, $S_c$ belonging
to the same simple factor group $H_l$
and the index $u_1$ runs over the $U(1)$ factor groups in $H$.
The second coefficient $\gp$ corresponding to the $G/H$ index $\a$
also takes a similar form:
$$
  \gp =
  \gampla {f_{bc}}^a \delta_\a^c
  + \tilde\gampla {d_{bc}}^a \delta_\a^c
  + \ttgamp \delta_b^a \delta_\a^{\nu_1} \ ,
\eqn\eqADD
$$
where $\gampla$ and $\tilde\gampla$
can be non-vanishing only when the
$H$-irreducible representation $[\a]$ to which the $G/H$ generator
$X_\a$ belongs happens to be the same representation as the $H_l$
generators, namely, adjoint under $H_l$ and singlet under the other
factor groups $H_k \  (k\neq l)$. The index $\nu_1$ in the last term
in \eqADD\ runs over all the $H$-singlet indices among the $G/H$
generators $X_\a$.

Eq.\delCCterm\  simply gives ${{\cal H}_{bi}}^a(\phi)$
in terms of ${G_{bj}}^a(\phi)$:
$$
  {{\cal H}_{bi}}^a (\phi) =
  - {G_{bj}}^a (\phi) {\cal W}_i^j (\phi).
\eqn{\SolcalH}
$$
This form says an interesting fact:
in \Vtrans, the terms ${G_{bi}}^a \partial_\mu\phi^i$
and ${{\cal H}_{bi}}^a{\cal V}_\mu^i$ are combined to yield
$$
  {G_{bi}}^a(\phi) \partial_\mu\phi^i
  + {{\cal H}_{bj}}^a (\phi) {\cal V}_\mu^j
  =
  {G_{bi}}^a
  \left({
    \partial_\mu \phi^i - {\cal V}_\mu^j {\cal W}_j^i (\phi)
  }\right) .
\eqn{\CovdelPhi}
$$
This is nothing but a $\Gglobal $-covariant derivative
since $\delta\phi^i={\cal W}_j^i$ is just an infinitesimal
$\Gglobal $ transformation by $T_j$.
In view of \Gbiaterm,
we therefore find that \CovdelPhi\  must equal
$$
\eqalign{
  \hbox{\rm \CovdelPhi\ }
&=
  {g_{bc}}^a {2\over i} \tr
  \Big[{
    S^c ({\partial_\mu \xi + i \xi {\cal V}_\mu}) \cdot \xi^{\dag}
  }\Big]
 + \gp {2\over i} \tr
  \Big[{
    X^\a ({\partial_\mu \xi + i \xi {\cal V}_\mu}) \cdot \xi^{\dag}
  }\Big]
\cr
&=
  {g_{bc}}^a \Apara_\mu^c (\phi) + \gp \Apm^\a (\phi)
}
\eqno\eq
$$
with $\Apara_\mu^a(\phi)$ defined in \Amupart\ and $\Apm^\a(\phi)$
defined similarly.

Eq.\CCVterm\  again says the constantness of ${H_{bc}}^a(\phi)$
so that we have from $\Hdiag $-covariance
$$
  {H_{bc}}^a(\phi) =
  h_l {f_{bc}}^a +
  \tilde{h}_l {d_{bc}}^a
  + \tilde{\tilde{h}}_{lu_1} \delta_b^a \delta_c^{u_1}
  \quad
  \left({ \equiv {h_{bc}}^a }\right) .
\eqno\eq
$$
The final eq.\CCcalVterm\  is easily seen to be satisfied
automatically using \CCdelPhiterm\  and \SolcalH.

At this stage, the $\deltaGam V_\mu^a$ in \Vtrans\
is already rather simplified:
$$
\eqalign{
  \deltaGam V_\mu^a
&=
  \alpha'_l \partial_\mu C^a +
  \left({
    {g_{bc}}^a \Apara_\mu^c (\phi)
     + \gp \Apm^\a (\phi) + {h_{bc}}^a V_\mu^c
  }\right)
  C^b
\cr
&=
  \alpha'_l \partial_\mu C^a +
  {\cal U}_{\mu,l}(\phi) C^a
  + {f_{bc}}^a B_\mu^c (\phi) C^b
  + {d_{bc}}^a \tilde{B}_\mu^c(\phi) C^b ,
}
\eqn{\DelDV}
$$
with notations
$$
\eqalign{
  {\cal U}_{\mu,l} (\phi)
&\equiv
  \tilde{\tilde{\gamma}}_{lu_1} \Apara_\mu^{u_1}(\phi)
  + \ttgamp \Apm^{\nu_1}(\phi)
  + \tilde{\tilde{h}}_{lu_1} V_\mu^{u_1} \ ,
\cr
  B_\mu^c (\phi)
&\equiv
  \gamma_l \Apara_\mu^c(\phi)
  + \gampla \Apm^{\a=c}(\phi) + h_l V_\mu^c \ ,
\cr
  \tilde{B}_\mu^c (\phi)
&\equiv
  \tilde{\gamma}_l \Apara_\mu^c (\phi)
  + \tilde\gampla \Apm^{\a=c}(\phi)
   + \tilde{h}_l V_\mu^c \ .
}
\eqn{\DefOfUBB}
$$
Since now the indices $a$, $b$ and $c$ in the second expression
in \DelDV\  for $\deltaGam V_\mu^a$
are all those belonging to the same simple factor group $H_l$,
it is much more convenient to switch to matrix notation
and rewrite \DelDV\  with \DefOfUBB\  into
$$
  \deltaGam V_\mu
  =
  \alpha' \partial_\mu C
  + {\cal U}_\mu C
  + i \left[{ B_\mu , C }\right]
  + \big\{{ \tilde{B}_\mu , C }\big\} ,
\eqno\eq
$$
$$
  B_\mu = \gamma \Apara_\mu +\gamma^\perp \Apm + h V_\mu ,
  \quad
  \tilde{B}_\mu = \tilde{\gamma} \Apara_\mu
   + \tilde\gamma^\perp \Apm + \tilde{h} V_\mu,
\eqno\eq
$$
where the matrices mean, for instance,
$$
  \alpha'C \equiv \sum_l \alpha'_l
  \sum_{a\in H_l} C^a S_a ,
  \quad
  h V_\mu \equiv \sum_l h_l \sum_{a\in H_l} V_\mu^a S_a,
$$
with summation taken only over simple factor groups $H_l$ in $H$.
Noting that $V_\mu^a - \Apara_\mu^a(\phi)$ and
$\Apm^\a(\phi)$ are $\Gglobal $-invariant
and $\Hlocal $-covariant, we find
$$
\eqalignno{
  \dBRS \Apara_\mu
&=
  \partial_\mu C - i \left[{ \Apara_\mu , C }\right] \ ,
\qquad
  \dBRS \Apm =
   - i \left[{ \Apm , C }\right] \ ,
&\eq
\cr
  \dBRS {\cal U}_{\mu,l}
&=
  \big({
    \tilde{\tilde{\gamma}}_{lu_1} + \tilde{\tilde{h}}_{lu_1}
  }\big) \partial_\mu C^{u_1}
  \equiv
  \partial_\mu C_{U(1),l} ,
&\eq
}
$$
and hence
$$
  \dBRS B_\mu = (\gamma+h) \partial_\mu C
  - i \left[{ B_\mu , C }\right] ,
  \quad
  \dBRS
  \left({
    \left[{ B_\mu , C }\right]
  }\right)
  =
  (\gamma+h) \partial_\mu ({ C^2 }) ,
\eqno\eq
$$
and so on.
Using these we calculate
$$
\eqalignno{
  \dBRS \left({ \deltaGam V_\mu }\right)
&=
  (\alpha'+\gamma+h) \partial_\mu
  \left({ i C^2 }\right)
  + \left({ \partial_\mu C_{U(1)} }\right) C
  + i {\cal U}_\mu C^2
& \
\cr
& \qquad\quad
  + \big({ \tilde{\gamma} + \tilde{h} }\big)
  \left[{ \partial_\mu C , C }\right]
  + 2i C \tilde{B}_\mu C ,
&\eqname{\dBdGV}
\cr
  \deltaGam \left({ \dBRS V_\mu }\right)
&=
  (\beta-\alpha') \partial_\mu
  \left({ i C^2 }\right)
  - 2 i {\cal U}_\mu C^2
  + \left[{ \beta V_\mu + B_\mu , C^2 }\right]
& \
\cr
& \qquad\quad
  - i \big\{{ \tilde{B}_\mu , C^2 }\big\}
  - 2 i C \tilde{B}_\mu C .
&\eqname{\dGdBV}
}
$$
These \dBdGV\  and \dGdBV\  should add up to cancel the last
$F_{[ab][cd]}$ term in \KlinV.
Since the latter term comes from
$F_{[ab][cd]} K_\mu^a K_\mu^b C^c C^d$
term in $\Gamma_6^{(n)}$,
the indices $a$ and $b$ must be of simple factor groups,
so that the last term contains neither $\Apara_\mu^{u_1}$
nor $\Apm^{\nu_1}$ nor $V_\mu^{u_1}$ of $U(1)$ factor group
or $H$-singlet indices.
Therefore the ${\cal U}_\mu C^2$ terms should cancel already
in \dBdGV$+$\dGdBV,
but it demands ${\cal U}_\mu$ itself vanish
(i.e., $\tilde{\tilde{\gamma}}=\tilde{\tilde{\gamma}}^\perp
=\tilde{\tilde{h}}=0$)  and so $C_{U(1)}=0$.
Moreover, the $F_{[ab][cd]}$ term in \KlinV\ contains only
$ V_\mu^b - \Apara_\mu^b$ of the simple factor groups $H_l$'s,
but no $\Apm^\a$ of the $G/H$ generators $X_\a$, so that the terms
containing $\Apm^\a$ in  \dBdGV$+$\dGdBV\ should also vanish;
that is, $\gamma^\perp = \tilde\gamma^\perp = 0$.  Further,
since the $F_{[ab][cd]}$ term in \KlinV\
does not contain $\partial_\mu C$ either,
the terms containing $\partial_\mu C$ should also vanish
in \dBdGV$+$\dGdBV.
This yields
$$
  \beta_l + \gamma_l + h_l = 0 ,
  \quad
  \tilde{\gamma}_l + \tilde{h}_l = 0 ,
\eqn{\Solbgh}
$$
so that \dBdGV$+$\dGdBV\  becomes, at this stage,
$$
  \left\{{ \dBRS , \deltaGam }\right\} V_\mu
  =
  - \gamma
  \left[{
    V_\mu - \Apara_\mu , C^2
  }\right]
  + i \tilde{\gamma}
  \left\{{
    V_\mu - \Apara_\mu , C^2
  }\right\} ,
\eqno\eq
$$
or, in terms of the original notation,
$$
  \left\{{ \dBRS , \deltaGam }\right\} V_\mu^a
  =
  {1\over2} \gamma_l {f_{bc}}^a
  \left({ V_\mu - \Apara_\mu }\right)^b
  \left({ C \times C }\right)^c
  - {1\over2} \tilde{\gamma}_l {d_{bc}}^a
  \left({ V_\mu - \Apara_\mu }\right)^b
  \left({ C \times C }\right)^c .
\eqno\eq
$$
The second term proportional to ${d_{bc}}^a$ is symmetric
under $a \leftrightarrow b$,
and cannot be canceled with the $a \leftrightarrow b$
anti-symmetric $F_{[ab][cd]}$ term in \KlinV.
So
$$
  \tilde{\gamma}_l = 0 ,
\eqn{\Solgtil}
$$
and then \KlinV\  turns out to give
$$
  F_{[ab][cd]} (\phi)
  =
  - {1\over4} { \gamma_l \over a_{\parallel l} }
  f_{abe} {f_{cd}}^e .
\eqno\eq
$$
Putting \Solbgh, \Solgtil, $\gamma^\perp=\tilde\gamma^\perp=0$
and ${\cal U}_{\mu,l}=0$ together into \DelDV, we finally obtain
$$
\eqalign{
  \deltaGam V_\mu^a
&=
  \alpha'_l \partial_\mu C^a
  + \beta_l \left({ V_\mu \times C }\right)^a
  + \gamma_l
  \left({
    \left({ V_\mu - \Apara_\mu }\right) \times C
  }\right)^a
\cr
&=
  \alpha_l \partial_\mu C^a
  + \beta_l \dBRS V_\mu^a
  + \gamma_l \dBRS \left({ V_\mu^a - \Apara_\mu^a }\right) ,
}
\eqno\eq
$$
with $\alpha_l \equiv \alpha'_l - \beta_l$.

We thus have finished solving the renormalization
equation \DimfourWTB\
and found the following collecting the results obtained above:
$$
\eqalignno{
  \left. \Gamma_4^{(n)} \right\vert_{\Klinear}
&=
  K_i
  \left[{
    {\mib C}^A
    \left({
      \hat{\mib W}_A F^i(\phi) - \hat{F} {\mib W}_A^i(\phi)
    }\right)
    + \beta_l C^a W_a^i(\phi)
  }\right]
& \
\cr
& \qquad
  + K_a^\mu
  \left[{
    \alpha_l \partial_\mu C^a + \beta_l \dBRS V_\mu^a
    + \gamma_l \dBRS
    \left({ V_\mu^a - \Apara_\mu^a (\phi) }\right)
  }\right]
& \
\cr
& \qquad
  + L_a
  \left({ \beta_l \dBRS C^a }\right) ,
&\eqname{\SolDimfour}
\cr
  \left. \Gamma_6^{(n)} \right\vert_{\Kquadratic}
&=
  - {1 \over 4f_\pi^2} {\gamma_l \over a_{\parallel l}}
  \left({ K_\mu \times K^\mu }\right)^a
  \left({ C \times C }\right)^a ,
&\eqname{\SolDimsix}
}
$$
where $F^i(\phi)$ is an arbitrary function and
$\alpha_l$, $\beta_l$, $\gamma_l$ are arbitrary parameters
dependent on the simple factor group $H_l$.

\section{ Solving Eq.\DimtwoWTB }

Now that we have determined the form of $\deltaGam$,
we can solve the renormalization equation \DimtwoWTB.
We calculate $\deltaGam S_2$ using \SolDimfour\  and
\SolDimsix\  as follows:
$$
\eqalignno{
  \deltaGam S_2
&=
  \left({ \deltaGam \phi^i }\right)
  { \delta S_2 \over \delta \phi^i }
  + \left({ \deltaGam V_\mu^a }\right)
  { \delta S_2 \over \delta V_\mu^a }
&\eqname{\delGStwo}
\cr
&=
  {\mib C}^A \big[{ \hat{\mib W}_A , \hat{F} }\big] S_2
&\eqname{\delGStwoA}
\cr
& \qquad
  + \beta_l
  \left({
    C^a W_a^i(\phi) {\delta \over \delta \phi^i}
    + D_\mu C^a {\delta \over \delta V_\mu^a}
  }\right)
  S_2
&\eqname{\delGStwoB}
\cr
& \qquad
  +
  \left[{
    \alpha_l \partial_\mu C^a
    + \gamma_l
    \left({
      \left({ V_\mu - \Apara_\mu }\right) \times C
    }\right)^a
  }\right]
  { \delta S_2 \over \delta V_\mu^a} .
&\eqname{\delGStwoC}
}
$$
First note that
$$
\eqalign{
  \big[{ \dBRS , \hat{F} }\big]
&=
  \left[{
    {\mib C}^A {\mib W}_A^i(\phi) {\delta \over \delta \phi^i}
    + D_\mu {\mib C}^A {\delta \over \delta \VA_\mu^A}
    - {1\over2} \left({ {\mib C} \times {\mib C} }\right)^A
    {\delta \over \delta {\mib C}^A}
    ,
    F^j(\phi) {\delta \over \delta \phi^j}
  }\right]
\cr
&=
  \big[{
    {\mib C}^A \hat{\mib W}_A , \hat{F}
  }\big] ,
}
\eqno\eq
$$
which together with $\dBRS S_2 = 0$ leads to
$$
  {\mib C}^A \big[{ \hat{\mib W}_A , \hat{F} }\big] S_2
  = \dBRS \left({ \hat{F} S_2 }\right) .
\eqno\eq
$$
Next we note that
$ \beta_l
\left({ C^a \hat{W}_a + D_\mu C^a \delta/\delta V_\mu^a }\right)$
appearing in \delGStwoB\  is a $H_l$-local gauge transformation
with angle $\theta^a = \beta_l C^a$,
and so \delGStwoB\  vanishes because of the $H_l$-gauge invariance
of $S_2$.  Thirdly, using the fact that
$\delta S_2 / \delta V_\mu^a =
a_{\parallel l} f_\pi^2 \left({ V_\mu^a - \Apara_\mu^a }\right)$
by \Lparallel, we see
$$
\eqalign{
  \hbox{\rm \delGStwoC }
&=
  \left({ \alpha_l \partial_\mu C^a }\right)
  \cdot
  a_{\parallel l} f_\pi^2
  \left({ V_\mu^a - \Apara_\mu^a }\right)
\cr
&=
  \dBRS
  \left({
    \alpha_l V_\mu^a \cdot a_{\parallel l} f_\pi^2
    \left({ V_\mu^a - \Apara_\mu^a }\right)
  }\right)
\cr
&=
  \dBRS
  \left({
    \alpha_l V_\mu^a {\delta \over \delta V_\mu^a} S_2
  }\right) .
}
\eqno\eq
$$
Thus we find that $\deltaGam S_2$ can be written in the form
$\dBRS (\star)$,
and the renormalization equation \DimtwoWTB\  becomes
$$
  \dBRS \Gamma_2^{(n)} + \dBRS
  \left({
    \hat{F} S_2 + \alpha_l V_\mu^a
    {\delta \over \delta V_\mu^a} S_2
  }\right)
  =0 .
\eqno\eq
$$
The general solution to this is clearly given by
$$
  \Gamma_2^{(n)} =
  A_{2\GI} [\phi,\VA] -
  \left({
    \hat{F} S_2 + \alpha_l V_\mu^a
    {\delta \over \delta V_\mu^a} S_2
  }\right) ,
\eqn{\SolDimtwo}
$$
with arbitrary gauge-invariant function $A_{2\GI}$ of dimension 2.

We have finished solving \DimtwoWTB.
It is now a trivial matter to check that our solutions
\SolDimtwo\  plus \SolDimfour\  and \SolDimsix\
are combined into a simple form
$$
  \Gamma_2^{(n)}
  + \left. \Gamma_4^{(n)} \right\vert_{\Klinear}
  + \left. \Gamma_6^{(n)} \right\vert_{\Kquadratic}
  =
  A_{2\GI} [\phi,\VA] - S \ast Y ,
\eqno\eq
$$
$$
  Y = \int d^4x
  \left[{
    K_i F^i (\phi)
    + \alpha_l K_a^\mu V_\mu^a
    + \beta_l L_a C^a
    + {\gamma_l \over 2 a_{\parallel l} f_\pi^2}
    f_{abc} K_a^\mu K_{b\mu} C^c
  }\right] ,
\eqno\eq
$$
aside from irrelevant terms which we are not discussing.
This just agrees with \Solution\  with \SolY\
which we wanted to prove.

\APPENDIX{B}{B.\  The proof of \SolOfRd }

We first note that \RelOfRd\  splits
into the following three types
according to whether the group indices $A$ and $B$
refer the $\Gglobal $ or $\Hlocal $ ones:
recalling $\hat{\mib W}_A = \big({ \hat{\cal W}_i , \hat{W}_a }\big)$
and $\hat{R}'_A = \big({ \hat{\cal R}'_i , \hat{R}'_a }\big)$
we have
$$
\eqalignno{
  \big[{ \hat{\cal W}_i , \hat{\cal R}'_j }\big]
  - \big[{ \hat{\cal W}_j , \hat{\cal R}'_i }\big]
&=
  {f_{ij}}^k \hat{\cal R}'_k ,
&\eqname{\RelOfRdA}
\cr
  \big[{ \hat{\cal W}_i , \hat{R}'_a }\big]
  - \big[{ \hat{W}_a , \hat{\cal R}'_i }\big]
&=
  0 ,
&\eqname{\RelOfRdB}
\cr
  \big[{ \hat{W}_a , \hat{R}'_b }\big]
  - \big[{ \hat{W}_b , \hat{R}'_a }\big]
&=
  {f_{ab}}^c \hat{R}'_c .
&\eqname{\RelOfRdC}
}
$$
We shall show below that the first set of equations, \RelOfRdA,
already gives enough information to determine the form
of the general solution $\hat{\cal R}_i$ as
$$
  \hat{\cal R}_i =
  \big[{ \hat{\cal W}_i , {}^\exists\hat{F} }\big] .
\eqn{\SolcalR}
$$
[ Recall that there are as many $\Gglobal $ generators
$\hat{\cal W}_i = -\partial/\partial\phi^i+\cdots$
as the variables $\phi^i$. ]
Assuming that \SolcalR\  is proved for $\hat{\cal R}_i$,
we first prove that $\hat{R}'_a$ is given by
$\big[{ \hat{W}_a , \hat{F} }\big]$
in terms of the same $\hat{F}$.
Eqs.\SolcalR, \RelOfRdB\  and the Jacobi identity
with the algebra $\big[{ \hat{\cal W}_i , \hat{W}_a }\big]=0$
lead to
$$
  \left[{
    \hat{\cal W}_i ,
    \Big({
      \hat{R}'_a -
      \big[{ \hat{W}_a , \hat{F} }\big]
    }\Big)
  }\right]
  = 0 .
\eqn{\ComRel}
$$
This does not immediately implies the desired eq.
$\hat{R}'_a = \big[{ \hat{W}_a , \hat{F} }\big]$,
since there are operators which commute with $\hat{\cal W}_i$'s.
But, note that $\hat{\cal W}_i$'s span a complete set of Lie algebra
generators of the group $G$ corresponding
to the {\it right}-multiplication.
Therefore, as is well-known,
the complete set of first order differential operators
which are commutative with all the $\hat{\cal W}_i$'s,
is given by the Lie algebra generators of $G$ corresponding to the
{\it left}-multiplication which we denote by $\hat{W}_i$.
Our $\Hlocal $ generators $\hat{W}_a$ are just a subset of
$\hat{W}_i$'s : $\hat{W}_{i=a}=\hat{W}_a$.
Thus, \ComRel\  generally says that the difference
$\hat{R}'_a - \big[{ \hat{W}_a , \hat{F} }\big]$
is given by a linear combination of the left-multiplication
generators:
$$
  \hat{R}'_a -
  \big[{ \hat{W}_a , \hat{F} }\big]
  =
  z_a^i \hat{W}_i
  \equiv
  \hat{Z}_a ,
\eqn{\LeftOp}
$$
with certain coefficients $z_a^i$.
We now use \RelOfRdC.
Since $\hat{R}'_a$ and clearly
$\big[{ \hat{W}_a , \hat{F} }\big]$ also satisfy \RelOfRdC,
so does $z_a^i \hat{W}_i \equiv \hat{Z}_a$ :
$$
  \big[{ \hat{W}_a , \hat{Z}_b }\big]
  - \big[{ \hat{W}_b , \hat{Z}_a }\big]
  =
  {f_{ab}}^c \hat{Z}_c .
\eqn{\RelOfZ}
$$
On the other hand, the $\Hdiag $-covariance implies
that both sides of \LeftOp\  have the same transformation law
under $\hat{W}_a+\hat{\cal W}_a$ [see \Hdiagtrans],
so that we have
$$
  \big[{ \hat{W}_a , \hat{Z}_b }\big]
  =
  \big[{ \hat{W}_a + \hat{\cal W}_a , \hat{Z}_b }\big]
  =
  {f_{ab}}^c \hat{Z}_c .
\eqno\eq
$$
Substituting this into \RelOfZ, we find
$$
  {f_{ab}}^c \hat{Z}_c = 0 ,
\eqno\eq
$$
which implies $\hat{Z}_a=0$ for the indices $a$ belonging
to any simple factor group $H_l$.
For the indices $a$ corresponding to $U(1)$-factor groups (if any),
however, $\hat{Z}_a = z_a^i \hat{W}_i$ still may not vanish.
But, again by the covariance under $\Hdiag $,
the coefficients $z_a^i$ can be nonvanishing only for
$\Hdiag $-singlet indices $i$.
Recall that we are discussing the $K_i$-term appearing
in $\Gamma_4^{(n)}$,
which now takes the form,
by \eqAXV, \SolcalR\  and \LeftOp
$$
\eqalign{
&
  K_i
  \left[{
    {\cal C}^j {\cal R}_j^i (\phi) + C^a R_a^i(\phi)
  }\right]
\cr
& \qquad\quad
  = K_i
  \left[{
    {\cal C}^j \big[{ \hat{\cal W}_j , \hat{F} }\big] \phi^i
    + C^a
    \Big({
      \beta_l \hat{W}_a
      + \big[{ \hat{W}_a , \hat{F} }\big]
      + z_a^j \hat{W}_j
    }\Big)
    \phi^i
  }\right] .
}
\eqn{\eqBX}
$$
Noting $z_a^j\hat{W}_j \phi^i = z_a^i + {\cal O}(\phi)$
(since $\hat{W}_j = \partial / \partial\phi^j + \cdots$ ),
we see that the last term contributes only when the index $i$
is of $\Hdiag $-singlet and the index $a$ is of $U(1)$ factor group.
But such term proportional to $K_i C^a$ with $\Hlocal $-singlet $i$
and $U(1)$ index $a$ has a particular property
in the original action $S$.
In $S$, $K_i$ is contained in the form
$K_i \left({ {\cal C}^j {\cal W}_j^i(\phi)
  + C^a W_a^i(\phi) }\right)$.
Since the relation $W_a^i(\phi)=-{\cal W}_a^i(\phi)$
holds for $\Hdiag $-singlet index $i$ and the $U(1)$ ghosts $C^a$
are free, the external (non-quantized) ghost ${\cal C}^{j=a}$
and the $U(1)$ ghosts $C^a$ are contained
in the $\Hdiag $-singlet $K_i$ term on the same footing
in the form
$K_i \left({ {\cal C}^a - C^a }\right) {\cal W}_a^j(\phi)$.
Therefore the $n$-th loop level effective action $\Gamma_4^{(n)}$
has to contain the $K_iC^a$ term with the same coefficient
function of $\phi$ as the $-K_i{\cal C}^a$ term.
(Recall that $\beta_l=0$ for the $U(1)$ index $a$.)
In view of the RHS of \eqBX, this implies $z_a^j=0$.
Thus we have proved
$\hat{R}'_a = \big[{ \hat{W}_a , \hat{F} }\big]$ for any $a$.

Remaining is the proof of \SolcalR.
We prove it by mathematical induction with respect to
the powers in $\phi$ of the solution ${\cal R}_i^{'j}(\phi)$
to \RelOfRdA.
Both ${\cal W}_i^j(\phi)$ and ${\cal R}_i^j(\phi)$
have terms of powers $\phi^n$ with $n=0,1,2,\cdots$.
We denote the $n$-th power terms by ${\cal W}_i^{(n)j}$
and ${\cal R}_i^{'(n)j}$, and the corresponding
operators ${\cal W}_i^{(n)j} {\partial \over \partial \phi^j}$
and ${\cal R}_i^{'(n)j} {\partial \over \partial \phi^j}$ by
$\hat{\cal W}_i^{(n)}$ and $\hat{\cal R}_i^{'(n)}$.
Then our claim is the following:
For any solution $\hat{\cal R}'_i$ to \RelOfRdA,
there exist a $(n+1)$-st order polynomial $F_{n+1}^i(\phi)$
in $\phi$ starting from a linear term
$$
\eqalignno{
  F_{n+1}^i(\phi)
&=
  {F^i}_j\phi^j + {F^i}_{j_1j_2} \phi^{j_1} \phi^{j_2}
  + \cdots
  + {F^i}_{j_1j_2\cdots j_{n+1}}
  \phi^{j_1} \phi^{j_2} \cdots \phi^{j_{n+1}}
& \
\cr
&\equiv
  F_{(1)}^i(\phi) + F_{(2)}^i(\phi) + \cdots
  + F_{(n+1)}^i(\phi) ,
&\eqname{\PolF}
}
$$
for which the commutator
$\big[{ \hat{\cal W}_i , \hat{F}_{n+1} }\big]$
( $\hat{F}_{n+1} \equiv F_{n+1}^i (\phi) \partial / \partial\phi^i$ )
gives the solution $\hat{\cal R}'_i$ correctly up to the $n$-th
power terms; namely
$$
  \hat{\cal R}'_i =
  \big[{ \hat{\cal W}_i , \hat{F}_{n+1} }\big]
  + {\cal O} (\phi^{n+1}) \times \partial / \partial\phi .
\eqn{\RelRF}
$$

{\bf Proof)} \
We can start the induction from $n=-1$.
Then $F_{n+1}^i(\phi)=F_0^i(\phi)$ of \PolF\
is zero by definition,
but $\hat{\cal R}'_i$ itself is
${\cal O}(\phi^0) \times \partial / \partial \phi$ and so
\RelRF\  is trivially true.
If \RelRF\  holds for a certain $n$, then the difference
$\hat{r}_i \equiv \hat{\cal R}'_i -
\big[{ \hat{\cal W}_i , \hat{F}_{n+1} }\big]$
starts from a $(n+1)$-st power term in $\phi$.
Since both $\hat{\cal R}'_i$ and
$\big[{ \hat{\cal W}_i , \hat{F}_{n+1} }\big]$
satisfy \RelOfRdA,
the difference $\hat{r}_i$ also satisfies it:
$$
  \big[{ \hat{\cal W}_i , \hat{r}_j }\big]
  - \big[{ \hat{\cal W}_j , \hat{r}_i }\big]
  =
  {f_{ij}}^k \hat{r}_k .
\eqno\eq
$$
Consider the $n$-th power terms on both sides of this equation.
Such terms exist only in the left-hand side and
come from the lowest power terms of both $\hat{\cal W}_i$
and $\hat{r}_i$.
So we have
$$
  \big[{ \hat{\cal W}_i^{(0)} , \hat{r}_j^{(n+1)} }\big]
  - \big[{ \hat{\cal W}_j^{(0)} , \hat{r}_i^{(n+1)} }\big]
  = 0 .
\eqno\eq
$$
Recalling $\hat{\cal W}_i^{(0)} = - \partial / \partial\phi^i$
\DefcalW\ and writing
$\hat{r}_i^{(n+1)} = r_i^{(n+1)j}(\phi) \partial / \partial\phi^j$,
this simply gives
$$
  {\partial \over \partial\phi^i}
  r_j^{(n+1)k}(\phi)
  - {\partial \over \partial\phi^j}
  r_i^{(n+1)k}(\phi)
  = 0 .
\eqno\eq
$$
But this is just an integrability condition and guarantees that
there exists a $(n+2)$-nd power homogeneous function
$F_{(n+2)}^i(\phi)$ such that
$$
  r_j^{(n+1)i} (\phi)
  =
  - {\partial \over \partial \phi^j}
  F_{(n+2)}^i (\phi)
  =
  \big[{ \hat{\cal W}_j^{(0)} , \hat{F}_{(n+2)} }\big]\phi^i .
$$
This implies that if we define $(n+2)$-nd order polynomial
$F_{n+2}^i(\phi)$ by
$$
  F_{n+2}^i (\phi) = F_{n+1}^i(\phi) + F_{(n+2)}^i(\phi) ,
$$
it satisfies
$$
  \hat{\cal R}'_i
  =
  \big[{ \hat{\cal W}_i , \hat{F}_{n+2} }\big]
  + {\cal O}(\phi^{n+2}) \times \partial / \partial\phi ,
$$
namely, \RelRF\  with $n$ raised by 1.
This finishes the proof.

\refout

\end